\documentclass[12pt,preprint]{aastex}

\slugcomment{\emph{To appear in:}\\ \textsc{The Astrophysical Journal}, 578, 2002 October 20}

\newcommand{\ud}{\mathrm{d}} 
\newcommand{\lsun}{\hbox{L$_\odot$}}
\newcommand{\kms}{\hbox{km s$^{-1}$}}

\shorttitle{Counterrotating core and black hole of IC~1459}
\shortauthors{Cappellari et al.}

\begin{document}

\title{The counterrotating core and the black hole mass of IC~1459\altaffilmark{1}}

\altaffiltext{1}{Based on observations with the NASA/ESA {\it Hubble Space Telescope} obtained at the Space Telescope Science Institute, which is operated by the AURA, Inc., under NASA contract NAS 5-26555. These
observations are associated with proposal no.\ 7352.}

\author{M.~Cappellari\altaffilmark{2,3},
E.~K.~Verolme\altaffilmark{3},
R.~P.~van~der~Marel\altaffilmark{4},
G.~A.~Verdoes~Kleijn\altaffilmark{3},
G.~D.~Illingworth\altaffilmark{5},
M.~Franx\altaffilmark{3},
C.~M.~Carollo\altaffilmark{6},
P.~T.~de~Zeeuw\altaffilmark{3}}

\altaffiltext{2}{European Space Agency external fellow}
\altaffiltext{3}{Leiden Observatory, Postbus 9513, 2300 RA Leiden, The Netherlands}
\altaffiltext{4}{Space Telescope Science Institute, 3700 San Martin Drive, Baltimore, MD 21218}
\altaffiltext{5}{Lick Observatory, University of California, Santa Cruz, CA 95064}
\altaffiltext{6}{Swiss Federal Institute of Technology ETH, Zurich, Switzerland}

\begin{abstract}

The E3 giant elliptical galaxy \objectname{IC~1459} is the
prototypical galaxy with a fast counterrotating stellar core. We
obtained one HST/STIS long-slit spectrum along the major axis of this
galaxy and CTIO spectra along five position angles. The signal-to-noise ($S/N$) of the
ground-based data is such that also the higher order Gauss-Hermite
moments ($h_3$--$h_6$) can be extracted reliably. We present
self-consistent three-integral axisymmetric models of the stellar
kinematics, obtained with Schwarzschild's numerical orbit
superposition method. The available data allow us to study the
dynamics of the kinematically decoupled core (KDC) in IC~1459 and we find it
consists of stars that are well-separated from the rest of the galaxy
in phase space.  In particular, our study indicates that the stars in
the KDC counterrotate in a disk on orbits that are close to circular.
We estimate that the KDC mass is $\approx0.5$\% of the total galaxy
mass or $\approx3\times10^9 M_\odot$.

We estimate the central black hole mass $M_\bullet$ of IC~1459 independently from both its stellar and its gaseous kinematics. Although both tracers rule out models without a central black hole, neither yields a particularly accurate determination of the black hole mass. The main problem for the stellar dynamical modeling is the fact that the modest $S/N$ of the STIS spectrum and the presence of strong gas emission lines preclude measuring the full line-of-sight velocity distribution (LOSVD) at HST resolution. The main problem for the gas dynamical modeling is that there is evidence that the gas motions are disturbed, possibly due to non-gravitational forces acting on the gas. These complications probably explain why we find rather discrepant BH masses with the different methods. The stellar kinematics suggest that $M_\bullet = (2.6 \pm 1.1) \times 10^9 M_\odot$ ($3\sigma$ error). The gas kinematics suggests that $M_\bullet \approx 3.5 \times 10^8 M_\odot$ if the gas is assumed to rotate at the circular velocity in a thin disk. If the observed velocity dispersion of the gas is assumed to be gravitational, then $M_\bullet$ could be as high as $\sim 1.0 \times 10^9 M_\odot$. These different estimates bracket the value $M_\bullet = (1.1\pm0.3)\times10^9 M_\odot$ predicted by the $M_\bullet$-$\sigma$ relation. It will be an important goal for future studies to attempt comparisons of black hole mass determinations from stellar and gaseous kinematics for other galaxies. This will assess the reliability of black hole mass determinations with either technique. This is essential if one wants to interpret the correlation between the BH mass and other global galaxy parameters (e.g. velocity dispersion) and in particular the scatter in these correlations (believed to be only $\sim 0.3$ dex).
\end{abstract}


\section{Introduction}

The E3 giant elliptical galaxy IC~1459 ($M_V\simeq-22.3$ [LEDA, see \citealt{pat97}], $D=30.3\pm4.0$ Mpc [\citealt{fer00}], $R_e\simeq40\farcs6$ [\citealt{bur93}]) is a member of a loose group consisting mainly of spirals. It shows clear signs of past interaction, evidenced by stellar ``spiral arms'' detected on deep photographs \citep{mal85} and shells at large radii \citep{for95b}. The nuclear region harbors a fast counterrotating stellar component \citep{fra88}, which cannot be identified as a separate structure in photometric observations. A blue LINER source (unresolved at HST resolution), surrounded by patchy dust absorption and by a gaseous disk, is found at the very center \citep*[hereafter VK00]{for95a,car97,ver00}.

VK00 measured the mass of the central supermassive black hole in IC~1459 by analyzing kinematical observations of the nuclear gas disk, obtained with the Faint Object Spectrograph (FOS) on board the Hubble Space Telescope (HST). They found a central black hole mass of $M_\bullet=(2-6)\times10^8 M_\odot$ (scaling their values to our adopted distance\footnote{The choice of the distance $D$ does not influence our conclusions but sets the scale of our models in physical units. Specifically, lengths and masses scale as $D$, while mass-to-light ratios scale as $D^{-1}$.
}). They also calculated isotropic two-integral axisymmetric dynamical models for ground-based major-axis stellar kinematics, resulting in a significantly larger mass estimate ($M_\bullet=[4-6]\times10^9 M_\odot$). For comparison, the $M_\bullet$-$\sigma$ relation \citep{fer00,geb00} as given by Tremaine et al.\ (2002) predicts a black hole mass of $(1.1\pm0.3)\times10^9 M_\odot$.

In this paper, we construct axisymmetric three-integral models for IC~1459, using Schwarz\-schild's orbit-superposition method \citep{rix97,van98,cre99b}. The models are constrained by HST/STIS stellar kinematics measurements along the major axis and deep, ground-based spectroscopic observations along five different position angles of the slit. We show that this data-set allows us to derive information on the nature of the kinematically decoupled component (KDC). We also
re-address the value for the mass of the nuclear BH in this galaxy.

This paper is organized as follows. We discuss the spectroscopic and photometric data in Section~2 and present the adopted dynamical model in Section~3. We discuss our results in Section~4 and summarize our conclusions  in Section~5.

\section{Data}
\subsection{HST spectroscopy}
\label{sec:hst_spec}

An HST spectrum of IC~1459 was obtained with STIS (GO 7352, PI: C.M. Carollo) along the galaxy major axis (as determined from the HST central isophotes), using the G430L grating with the 52\arcsec$\times$0\farcs1 slit. Four exposures, slightly shifted in the spatial direction, were obtained for a total of 9812 seconds. The availability of four shifted images allowed for a good removal of the numerous hot and bad pixels, together with cosmic rays, during the coaddition.

The extraction of the stellar kinematics was performed using the pixel-space Line-Of-Sight Velocity Distribution (LOSVD) fitting method of \citet{van94b}. In pixel space, the masking of the prominent gas emission lines during the fit becomes straightforward. Specifically, a K3 III template star, observed with the 0\farcs2 STIS slit (GO 7388, PI: D. Richstone; no acceptable stellar template was available for the 0\farcs1 slit), was convolved with a parametrized LOSVD. The parameters of the LOSVD were optimized to the lowest $\chi^2$ by direct comparison with the observed spectrum. The galaxy spectrum was rebinned spatially before fitting, to an average S/N$\approx$20 per \AA. The error images produced by the STScI pipeline were used to weight the measurements and estimate the errors in the fit. Our IDL\footnote{http://www.rsinc.com} fitting procedure differs from that of \citet{van94b} in the following aspects:
\begin{enumerate}
\item We used the elegant {\tt DIRECT} deterministic global optimization algorithm \citep*{jon93} to perform the $\chi^2$ optimization for the nonlinear variables $V$ and $\sigma$. This method is \emph{guaranteed} to find the global minimum of a (possibly non-smooth) nonlinear function inside a finite box, even in the presence of multiple minima. It can be effectively used in many non-smooth real-world optimization problems with a small number of variables ($N\la10$).

\item We performed a $3\sigma$ clipped fit to make the fit stable against outliers (e.g. residual bad pixels, unmasked gas emission lines or template mismatch). In practice, once the global minimum is found, the galaxy pixels which deviate more than $3\sigma$ from the best fitting template spectrum are added to the list of masked pixels. The global optimization is repeated from scratch on the new set of pixels. This process is iterated until the set of bad pixels does not change anymore.

\item The errors on the $V$ and $\sigma$ fitting parameters are determined from the confidence levels for a $\chi^2$ distribution with two degrees of freedom.
\end{enumerate}

The resulting dispersion measurements $\sigma_\mathrm{fit}$ were corrected by approximating the line-spread-function (LSF) by a Gaussian, for both the galaxy and the template star. The template was observed with an effective instrumental dispersion $\sigma_{{\rm LSF},t}\approx100$ \kms\ \citep{lei01}, while the galaxy was observed with  $\sigma_{{\rm LSF},g}\approx167$ \kms. The effective instrumental dispersion for the template star is {\em smaller} than that of the galaxy, even though the star was observed with a {\em larger} slit. The reason for this is that the STIS PSF falls completely within the 0\farcs2 slit that was used to observe the template, so that the template LSF is determined essentially by the PSF size. The corrected dispersion $\sigma$ was obtained from
\begin{equation}
    \sigma^2 = \sigma^2_{\rm fit} - \sigma^2_{{\rm LSF},g} + \sigma^2_{{\rm LSF},t}.
\end{equation}
This correction is of the same order of magnitude as the measurement errors. No kinematic measurements could be extracted from the two central rows (0\farcs05 from the galaxy center) of the STIS spectrum due to the sharp decrease of the absorption line S/N of the stellar component, caused by the rise of a featureless non-thermal continuum from the central source. The non-thermal nature of the continuum is indicated by the sharp decrease of the line-strength parameter $\gamma$ towards the center.

The measured kinematical profiles are presented in Table~\ref{tab:stis_kin} and  in Figure~\ref{fig:plot_fit_stis}.

\begin{deluxetable}{rrrrr}
\tablewidth{0pc}
\tablecaption{IC~1459, PA=34$^\circ$, STIS stellar kinematics.\label{tab:stis_kin}}
\tablehead{
\colhead{R} & \colhead{$V$} & \colhead{$\Delta V$} & \colhead{$\sigma$} & \colhead{$\Delta \sigma$}  \\
\colhead{(\arcsec)} & \colhead{(\kms)}  & \colhead{(\kms)} & \colhead{(\kms)} & \colhead{(\kms)}
}
\startdata
 -1.50  & -146   &   20  &    296 &   22 \\
 -1.03  & -146   &   20  &    269 &   20 \\
 -0.73  & -109   &   20  &    309 &   22 \\
 -0.48  & -161   &   22  &    295 &   22 \\
 -0.25  & -72    &   30  &    349 &   24 \\
 -0.10  & -87    &   38  &    324 &   42 \\
  0.10  &  83    &   42  &    359 &   42 \\
  0.22  &  91    &   24  &    315 &   26 \\
  0.40  &  128   &   20  &    309 &   20 \\
  0.62  &  128   &   22  &    319 &   22 \\
  0.92  &  150   &   22  &    350 &   26 \\
  1.37  &  76    &   22  &    298 &   22 \\
\enddata
\end{deluxetable}

\subsection{Ground-based spectroscopy}

\begin{figure}
\epsscale{0.55}
\plotone{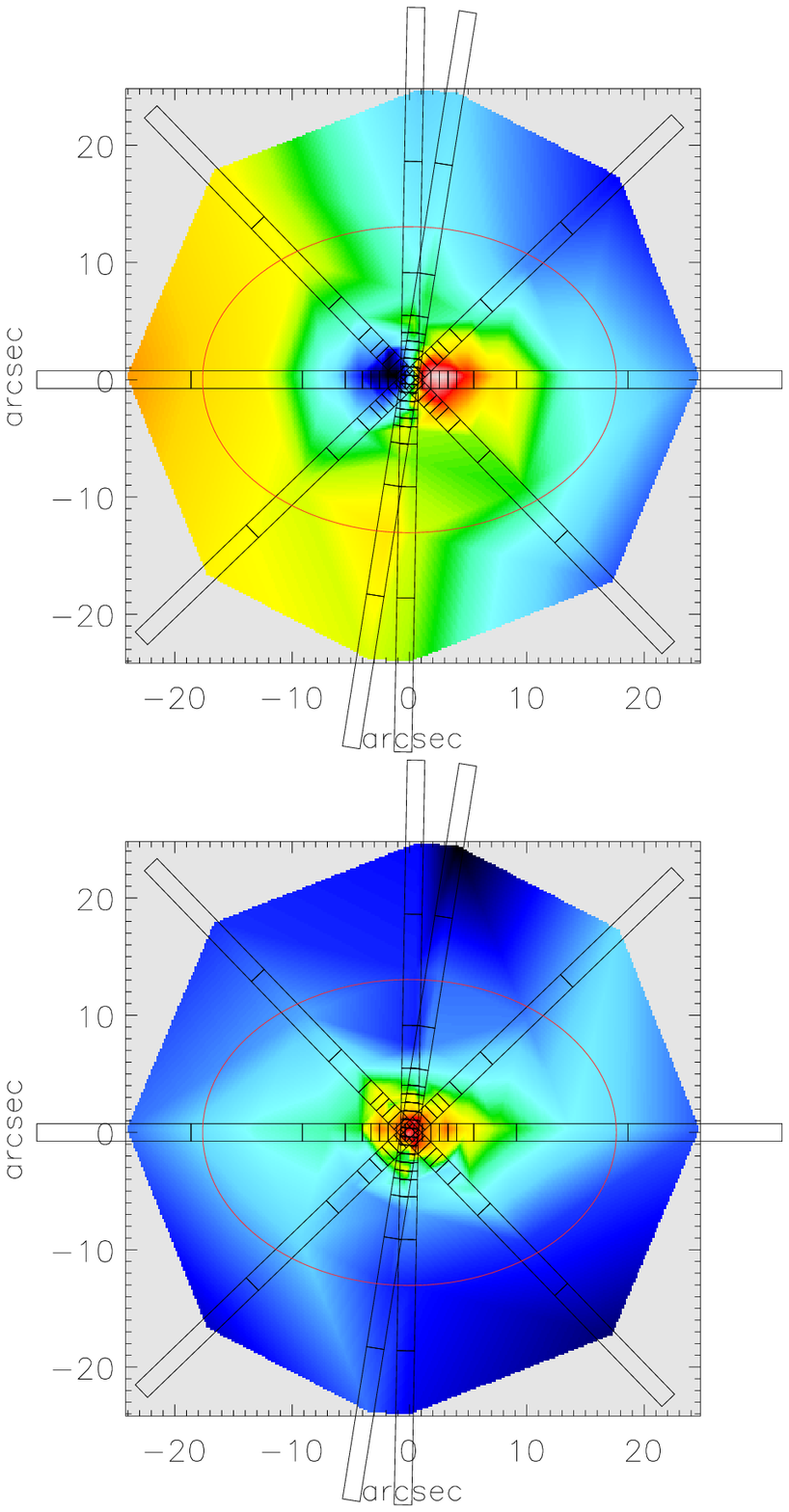}
\caption{Interpolated velocity field of IC~1459. \emph{Upper Panel:} mean velocity field, obtained by linear interpolation from the measured values at the aperture positions. The central counterrotating component is clearly visible. A representative galaxy isophote, the five slit positions and the regions along the slits that were averaged are also shown. \emph{Lower Panel:} same as in the upper panel for the observed velocity dispersion field. \label{fig:ic1459_kinematics}}
\end{figure}

We also analyzed a set of ground-based spectroscopic observations taken with the CTIO 4m telescope in November 1988. These observations were made along five different position angles with a 1\farcs5 wide slit, using a CCD with a 0\farcs73 pixel size. The seeing PSF was $\approx1\farcs5$ FWHM during the observations. The major axis, two $\pm45^\circ$ intermediate axes and two position angles close to the minor axis were observed. The location of the different slit positions is shown in Figure~\ref{fig:ic1459_kinematics}.  The major axis data were presented previously in \citet{van93}. The mean velocity $V$, the velocity dispersion $\sigma$ and the higher order Gauss-Hermite moments ($h_3$--$h_6$) were extracted from the high signal-to-noise spectra by the method described in that paper. The data are presented in Table~\ref{tab:kin_ctio}.

\clearpage
\begin{deluxetable}{rrrrrrrrrrrrr}
\tablecolumns{13}
\tablewidth{0pc}
\rotate
\tablecaption{IC~1459, CTIO stellar kinematics and velocity profiles shape.\label{tab:kin_ctio}}
\tablehead{
\colhead{R} & \colhead{$V$} & \colhead{$\Delta V$} & \colhead{$\sigma$} & \colhead{$\Delta \sigma$} & \colhead{$h_3$} & \colhead{$\Delta h_3$} & \colhead {$h_4$} & \colhead{$\Delta h_4$} & \colhead{$h_5$} & \colhead{$\Delta h_5$} & \colhead{$h_6$} & \colhead{$\Delta h_6$} \\
\colhead{(\arcsec)} & \colhead{(\kms)}  & \colhead{(\kms)} & \colhead{(\kms)} & \colhead{(\kms)} & \colhead{} & \colhead{} & \colhead{} & \colhead{} & \colhead{} & \colhead{} & \colhead{} & \colhead{}
}
\startdata
\cutinhead{PA=39$^\circ$}
 -25.2 &   36.6 &    8.9 &  259.1 &    8.8 & -0.008 &  0.031 & -0.034 &  0.030 & -0.017 &  0.033 &  0.077 &  0.031 \\
 -13.9 &   20.1 &    6.5 &  286.8 &    6.3 &  0.057 &  0.018 & -0.027 &  0.019 & -0.053 &  0.021 &  0.056 &  0.021 \\
  -7.3 &  -20.6 &    6.6 &  292.8 &    6.3 &  0.058 &  0.018 & -0.003 &  0.019 & -0.054 &  0.021 &  0.038 &  0.021 \\
  -4.7 &  -51.5 &    7.5 &  297.6 &    7.4 &  0.088 &  0.020 & -0.009 &  0.021 & -0.062 &  0.023 &  0.038 &  0.023 \\
  -3.7 &  -75.0 &    9.9 &  326.9 &    9.5 &  0.076 &  0.022 & -0.019 &  0.023 & -0.057 &  0.025 &  0.012 &  0.025 \\
  -2.9 &  -74.2 &    9.2 &  338.3 &    8.9 &  0.077 &  0.019 & -0.006 &  0.020 & -0.102 &  0.022 &  0.021 &  0.022 \\
  -2.2 &  -84.5 &    7.7 &  333.0 &    7.3 &  0.085 &  0.016 &  0.008 &  0.017 & -0.062 &  0.019 &  0.023 &  0.019 \\
  -1.5 &  -85.5 &    6.5 &  339.9 &    6.5 &  0.079 &  0.014 &  0.016 &  0.014 & -0.093 &  0.016 &  0.011 &  0.016 \\
  -0.7 &  -54.8 &    5.7 &  344.3 &   12.0 &  0.049 &  0.012 &  0.017 &  0.012 & -0.087 &  0.013 &  0.021 &  0.013 \\
   0.0 &   -4.7 &    5.4 &  355.8 &   12.0 & -0.030 &  0.011 & -0.012 &  0.011 & -0.041 &  0.013 &  0.033 &  0.012 \\
   0.7 &   58.0 &    5.6 &  342.6 &   12.0 & -0.068 &  0.012 & -0.041 &  0.012 &  0.003 &  0.014 &  0.068 &  0.014 \\
   1.5 &   79.9 &    6.5 &  332.3 &    6.2 & -0.119 &  0.015 &  0.003 &  0.015 &  0.008 &  0.016 &  0.048 &  0.017 \\
   2.2 &   84.3 &    7.5 &  329.1 &    7.3 & -0.158 &  0.017 & -0.004 &  0.018 & -0.006 &  0.020 &  0.029 &  0.020 \\
   2.9 &   80.7 &    9.1 &  339.9 &    9.1 & -0.148 &  0.021 & -0.013 &  0.021 &  0.032 &  0.023 &  0.082 &  0.024 \\
   3.7 &   75.7 &   10.0 &  333.2 &    9.4 & -0.115 &  0.022 &  0.003 &  0.022 & -0.034 &  0.025 & -0.001 &  0.025 \\
   4.7 &   54.1 &    8.2 &  319.2 &    7.7 & -0.081 &  0.019 & -0.036 &  0.020 & -0.015 &  0.022 &  0.074 &  0.023 \\
   7.3 &   26.7 &    6.6 &  303.9 &    6.1 & -0.072 &  0.017 & -0.045 &  0.018 &  0.017 &  0.020 &  0.080 &  0.020 \\
  13.9 &  -12.9 &    6.4 &  283.3 &    5.9 &  0.009 &  0.018 & -0.044 &  0.019 &  0.002 &  0.021 &  0.057 &  0.020 \\
  25.2 &  -38.7 &    8.8 &  255.5 &    8.8 &  0.050 &  0.030 & -0.037 &  0.030 & -0.058 &  0.033 &  0.045 &  0.031 \\
\cutinhead{PA=83$^\circ$}
 -25.2 &   18.3 &   10.4 &  238.1 &   10.7 & -0.064 &  0.041 & -0.013 &  0.038 & -0.016 &  0.042 &  0.054 &  0.040 \\
 -13.9 &   10.9 &    7.4 &  273.2 &    6.9 & -0.027 &  0.022 & -0.019 &  0.023 & -0.002 &  0.025 &  0.044 &  0.024 \\
  -7.3 &   -9.9 &    7.0 &  268.0 &    7.0 &  0.051 &  0.022 &  0.036 &  0.022 & -0.060 &  0.025 & -0.004 &  0.024 \\
  -4.7 &  -31.5 &    7.6 &  272.9 &    7.7 &  0.028 &  0.024 &  0.028 &  0.025 & -0.034 &  0.027 &  0.016 &  0.026 \\
  -3.7 &  -38.3 &   10.3 &  300.6 &   10.3 &  0.068 &  0.026 &  0.066 &  0.028 & -0.046 &  0.031 & -0.051 &  0.031 \\
  -2.9 &  -41.3 &    8.7 &  304.9 &    8.5 &  0.065 &  0.022 &  0.042 &  0.023 & -0.029 &  0.026 & -0.003 &  0.026 \\
  -2.2 &  -44.4 &    7.8 &  317.4 &    7.4 &  0.062 &  0.018 & -0.011 &  0.019 & -0.078 &  0.021 &  0.013 &  0.021 \\
  -1.5 &  -50.8 &    6.6 &  336.7 &    6.6 &  0.079 &  0.014 &  0.007 &  0.015 & -0.083 &  0.016 &  0.010 &  0.016 \\
  -0.7 &  -32.4 &    6.2 &  370.2 &   12.0 &  0.027 &  0.012 &  0.006 &  0.012 & -0.063 &  0.013 &  0.019 &  0.013 \\
   0.0 &    4.7 &    6.0 &  370.0 &   12.0 & -0.036 &  0.011 & -0.006 &  0.011 & -0.061 &  0.013 &  0.027 &  0.012 \\
   0.7 &   45.9 &    6.1 &  347.7 &   12.0 & -0.053 &  0.013 & -0.024 &  0.013 & -0.015 &  0.014 &  0.055 &  0.014 \\
   1.5 &   53.1 &    6.9 &  338.6 &    6.7 & -0.049 &  0.015 &  0.004 &  0.015 &  0.017 &  0.017 &  0.026 &  0.017 \\
   2.2 &   48.4 &    8.0 &  325.7 &    8.0 & -0.074 &  0.018 &  0.006 &  0.019 & -0.031 &  0.021 &  0.024 &  0.022 \\
   2.9 &   31.1 &    8.8 &  308.7 &    8.7 & -0.064 &  0.022 & -0.007 &  0.023 & -0.033 &  0.026 &  0.053 &  0.026 \\
   3.7 &   17.7 &   10.6 &  309.4 &   10.2 & -0.066 &  0.026 & -0.016 &  0.028 & -0.032 &  0.030 &  0.054 &  0.031 \\
   4.7 &   14.3 &    8.3 &  295.6 &    8.2 & -0.017 &  0.022 & -0.005 &  0.023 & -0.028 &  0.026 &  0.049 &  0.026 \\
   7.3 &   -7.8 &    7.1 &  279.6 &    7.0 & -0.034 &  0.021 & -0.005 &  0.022 & -0.028 &  0.024 &  0.034 &  0.023 \\
  13.9 &  -37.8 &    6.8 &  260.7 &    6.8 &  0.034 &  0.023 &  0.015 &  0.023 & -0.053 &  0.025 &  0.011 &  0.024 \\
  25.2 &  -56.4 &   10.6 &  273.3 &   10.3 &  0.041 &  0.032 & -0.027 &  0.033 & -0.054 &  0.037 &  0.066 &  0.036 \\
\cutinhead{PA=120$^\circ$}
 -25.2 &   11.0 &   12.2 &  257.8 &   11.7 &  0.034 &  0.040 & -0.069 &  0.040 &  0.028 &  0.045 &  0.077 &  0.042 \\
 -13.9 &   23.0 &    7.4 &  248.8 &    7.6 &  0.053 &  0.026 &  0.014 &  0.026 & -0.012 &  0.029 & -0.009 &  0.027 \\
  -7.3 &    5.0 &    6.9 &  259.3 &    7.1 &  0.031 &  0.023 &  0.031 &  0.023 & -0.014 &  0.026 &  0.007 &  0.024 \\
  -4.7 &   13.2 &    9.1 &  294.9 &    9.0 &  0.052 &  0.024 &  0.058 &  0.026 & -0.073 &  0.029 & -0.036 &  0.028 \\
  -3.6 &    7.6 &   11.1 &  311.5 &   10.6 &  0.008 &  0.026 & -0.001 &  0.028 & -0.045 &  0.031 &  0.009 &  0.031 \\
  -2.9 &   -9.7 &    9.4 &  294.8 &    9.2 &  0.036 &  0.025 &  0.042 &  0.026 & -0.090 &  0.029 &  0.002 &  0.029 \\
  -2.2 &   -2.9 &    7.8 &  294.8 &    7.5 &  0.053 &  0.020 &  0.021 &  0.022 & -0.061 &  0.024 &  0.011 &  0.024 \\
  -1.5 &    2.2 &    6.8 &  322.2 &    6.3 & -0.007 &  0.015 &  0.001 &  0.016 & -0.035 &  0.018 &  0.001 &  0.018 \\
  -0.7 &  -10.0 &    5.9 &  359.3 &   12.0 & -0.013 &  0.012 &  0.002 &  0.012 & -0.033 &  0.014 &  0.004 &  0.013 \\
   0.0 &   -3.8 &    5.9 &  380.0 &   12.0 & -0.001 &  0.011 & -0.015 &  0.011 & -0.034 &  0.012 &  0.023 &  0.012 \\
   0.7 &   17.1 &    6.3 &  363.4 &   12.0 & -0.011 &  0.012 & -0.006 &  0.012 & -0.022 &  0.014 &  0.035 &  0.013 \\
   1.5 &    1.3 &    6.7 &  331.5 &    6.8 & -0.011 &  0.015 &  0.017 &  0.016 & -0.025 &  0.018 &  0.014 &  0.018 \\
   2.2 &   -1.5 &    8.1 &  310.3 &    7.9 &  0.029 &  0.019 &  0.007 &  0.021 & -0.030 &  0.023 &  0.046 &  0.023 \\
   2.9 &   -4.9 &    9.5 &  299.3 &    9.6 &  0.011 &  0.024 &  0.050 &  0.026 & -0.036 &  0.029 &  0.001 &  0.029 \\
   3.6 &   -4.5 &   10.7 &  291.1 &   10.4 & -0.006 &  0.028 &  0.021 &  0.030 & -0.024 &  0.034 & -0.002 &  0.033 \\
   4.7 &   -4.8 &    8.6 &  288.8 &    8.4 &  0.008 &  0.023 & -0.025 &  0.025 &  0.012 &  0.028 &  0.049 &  0.027 \\
   7.3 &   -1.9 &    6.9 &  262.8 &    7.1 & -0.012 &  0.023 &  0.035 &  0.023 &  0.003 &  0.026 &  0.006 &  0.025 \\
  13.9 &  -15.9 &    7.6 &  257.8 &    7.9 & -0.023 &  0.026 & -0.001 &  0.026 &  0.040 &  0.029 &  0.062 &  0.028 \\
  25.2 &  -26.0 &   10.9 &  217.3 &   11.7 &  0.087 &  0.047 &  0.027 &  0.044 &  0.016 &  0.049 & -0.026 &  0.048 \\
\cutinhead{PA=128$^\circ$}
 -25.2 &    5.6 &   11.6 &  242.5 &   11.7 & -0.006 &  0.044 & -0.046 &  0.042 &  0.012 &  0.047 &  0.118 &  0.045 \\
 -13.9 &   10.6 &    7.2 &  247.3 &    7.5 &  0.030 &  0.027 &  0.003 &  0.026 &  0.007 &  0.029 &  0.044 &  0.027 \\
  -7.3 &    4.1 &    7.1 &  266.9 &    7.0 &  0.010 &  0.022 & -0.006 &  0.023 & -0.059 &  0.025 &  0.015 &  0.024 \\
  -4.7 &    6.1 &    8.2 &  277.9 &    8.2 & -0.060 &  0.024 &  0.012 &  0.025 & -0.049 &  0.028 &  0.011 &  0.027 \\
  -3.6 &   -4.1 &   10.1 &  284.7 &    9.8 &  0.011 &  0.029 &  0.023 &  0.030 & -0.031 &  0.033 &  0.018 &  0.033 \\
  -2.9 &   -0.0 &    8.7 &  291.3 &    8.2 & -0.028 &  0.023 & -0.034 &  0.025 & -0.080 &  0.027 &  0.053 &  0.027 \\
  -2.2 &   25.2 &    8.6 &  327.7 &    8.3 & -0.028 &  0.019 & -0.016 &  0.020 & -0.056 &  0.022 &  0.036 &  0.023 \\
  -1.5 &   -2.3 &    7.5 &  351.2 &    7.1 & -0.015 &  0.015 & -0.007 &  0.015 & -0.028 &  0.017 &  0.033 &  0.017 \\
  -0.7 &   -0.4 &    6.2 &  365.1 &   12.0 & -0.013 &  0.012 & -0.023 &  0.012 & -0.038 &  0.014 &  0.036 &  0.013 \\
   0.0 &   -1.3 &    6.1 &  371.4 &   12.0 & -0.027 &  0.011 & -0.022 &  0.011 & -0.057 &  0.013 &  0.052 &  0.012 \\
   0.7 &  -10.8 &    6.2 &  356.8 &   12.0 & -0.013 &  0.013 & -0.028 &  0.013 & -0.045 &  0.014 &  0.070 &  0.014 \\
   1.5 &   -1.8 &    6.9 &  333.4 &    6.7 &  0.002 &  0.015 & -0.004 &  0.016 & -0.040 &  0.018 &  0.030 &  0.018 \\
   2.2 &   -9.2 &    8.3 &  319.5 &    8.1 &  0.002 &  0.019 &  0.011 &  0.020 & -0.036 &  0.022 &  0.028 &  0.023 \\
   2.9 &   -7.4 &    9.3 &  312.2 &    9.1 &  0.001 &  0.023 &  0.016 &  0.024 & -0.061 &  0.027 &  0.020 &  0.027 \\
   3.6 &    1.5 &   10.7 &  296.3 &   10.7 & -0.008 &  0.028 & -0.009 &  0.030 & -0.080 &  0.033 &  0.051 &  0.033 \\
   4.7 &    5.0 &    8.6 &  289.5 &    8.5 &  0.008 &  0.024 &  0.019 &  0.025 & -0.128 &  0.028 &  0.010 &  0.027 \\
   7.3 &  -10.4 &    6.6 &  248.6 &    6.4 &  0.027 &  0.023 & -0.021 &  0.022 & -0.059 &  0.025 &  0.033 &  0.024 \\
  13.9 &  -19.3 &    7.1 &  248.9 &    7.4 & -0.002 &  0.026 &  0.011 &  0.025 & -0.022 &  0.028 &  0.031 &  0.027 \\
  25.2 &  -29.1 &   11.9 &  245.1 &   11.2 &  0.065 &  0.040 & -0.030 &  0.039 & -0.010 &  0.044 & -0.025 &  0.041 \\
\cutinhead{PA=173$^\circ$}
 -25.2 &  -45.5 &   10.5 &  229.5 &   10.9 &  0.046 &  0.042 &  0.003 &  0.040 &  0.004 &  0.044 &  0.039 &  0.043 \\
 -13.9 &    0.1 &    6.8 &  250.6 &    6.6 &  0.017 &  0.024 & -0.031 &  0.023 & -0.004 &  0.026 &  0.058 &  0.025 \\
  -7.3 &    5.0 &    7.3 &  290.0 &    7.2 & -0.029 &  0.020 &  0.012 &  0.021 &  0.006 &  0.024 &  0.037 &  0.023 \\
  -4.7 &   33.6 &    7.6 &  275.7 &    7.5 & -0.072 &  0.023 & -0.012 &  0.023 & -0.028 &  0.026 &  0.034 &  0.025 \\
  -3.7 &   47.9 &    9.7 &  308.7 &    9.4 & -0.088 &  0.024 & -0.006 &  0.026 &  0.040 &  0.028 &  0.041 &  0.029 \\
  -2.9 &   59.0 &    8.8 &  311.4 &    8.4 & -0.089 &  0.022 & -0.017 &  0.023 &  0.009 &  0.025 &  0.045 &  0.026 \\
  -2.2 &   57.7 &    8.0 &  329.5 &    7.8 & -0.063 &  0.018 & -0.007 &  0.019 &  0.001 &  0.021 &  0.039 &  0.021 \\
  -1.5 &   67.4 &    6.7 &  336.3 &    6.8 & -0.063 &  0.015 &  0.013 &  0.015 & -0.014 &  0.017 &  0.025 &  0.017 \\
  -0.7 &   41.3 &    6.0 &  349.1 &   12.0 & -0.060 &  0.012 & -0.024 &  0.012 & -0.021 &  0.014 &  0.053 &  0.014 \\
   0.0 &   -3.9 &    5.8 &  357.8 &   12.0 & -0.022 &  0.011 & -0.019 &  0.011 & -0.017 &  0.013 &  0.037 &  0.013 \\
   0.7 &  -40.4 &    5.7 &  344.5 &   12.0 &  0.006 &  0.012 & -0.001 &  0.012 & -0.045 &  0.014 &  0.024 &  0.014 \\
   1.5 &  -61.3 &    6.7 &  336.2 &    6.6 &  0.046 &  0.014 &  0.003 &  0.015 & -0.066 &  0.016 &  0.005 &  0.016 \\
   2.2 &  -61.9 &    7.9 &  321.1 &    7.6 &  0.067 &  0.017 &  0.032 &  0.018 & -0.033 &  0.021 & -0.004 &  0.021 \\
   2.9 &  -71.9 &    8.2 &  299.5 &    8.2 &  0.054 &  0.021 &  0.006 &  0.022 & -0.014 &  0.025 &  0.027 &  0.025 \\
   3.7 &  -34.2 &   10.4 &  314.4 &   10.0 &  0.003 &  0.024 &  0.027 &  0.026 & -0.030 &  0.029 &  0.004 &  0.029 \\
   4.7 &  -25.4 &    9.1 &  312.5 &    9.0 &  0.006 &  0.022 &  0.042 &  0.023 & -0.062 &  0.026 & -0.006 &  0.026 \\
   7.3 &   -9.1 &    7.0 &  286.3 &    7.1 & -0.002 &  0.020 &  0.021 &  0.021 & -0.035 &  0.023 &  0.023 &  0.022 \\
  13.9 &   12.8 &    7.1 &  265.9 &    7.6 &  0.003 &  0.023 &  0.050 &  0.024 &  0.007 &  0.026 & -0.013 &  0.025 \\
  25.2 &   19.0 &   10.3 &  250.5 &   10.2 & -0.033 &  0.037 & -0.023 &  0.036 &  0.111 &  0.040 &  0.020 &  0.038
\enddata
\end{deluxetable}
\clearpage

\subsection{HST imaging: correction of dust absorption\label{sec:dust}}

For our dynamical modeling, we need to derive an $I$-band surface density model from the available WFPC2/F814W (460 s) Hubble Space Telescope image of IC~1459. A complication is that this galaxy contains gas and dust in its central regions, which has to be corrected for in order to obtain an accurate determination of the central stellar surface brightness. We derive this correction in a manner that is similar to \citet[hereafter C97]{car97} and assume
\begin{enumerate}
\item the intrinsic color of the galaxy pixels that are affected by dust is the same as that of the surrounding pixels;
\item the dust is a screen in front of the stellar emission;
\item the Galactic extinction law \citep*{car89} applies.
\end{enumerate}
In this process, we use the available WFPC2/F555W (1000 s) frame in addition to the WFPC2/F814W image.

\begin{figure}
\epsscale{0.55}
\plotone{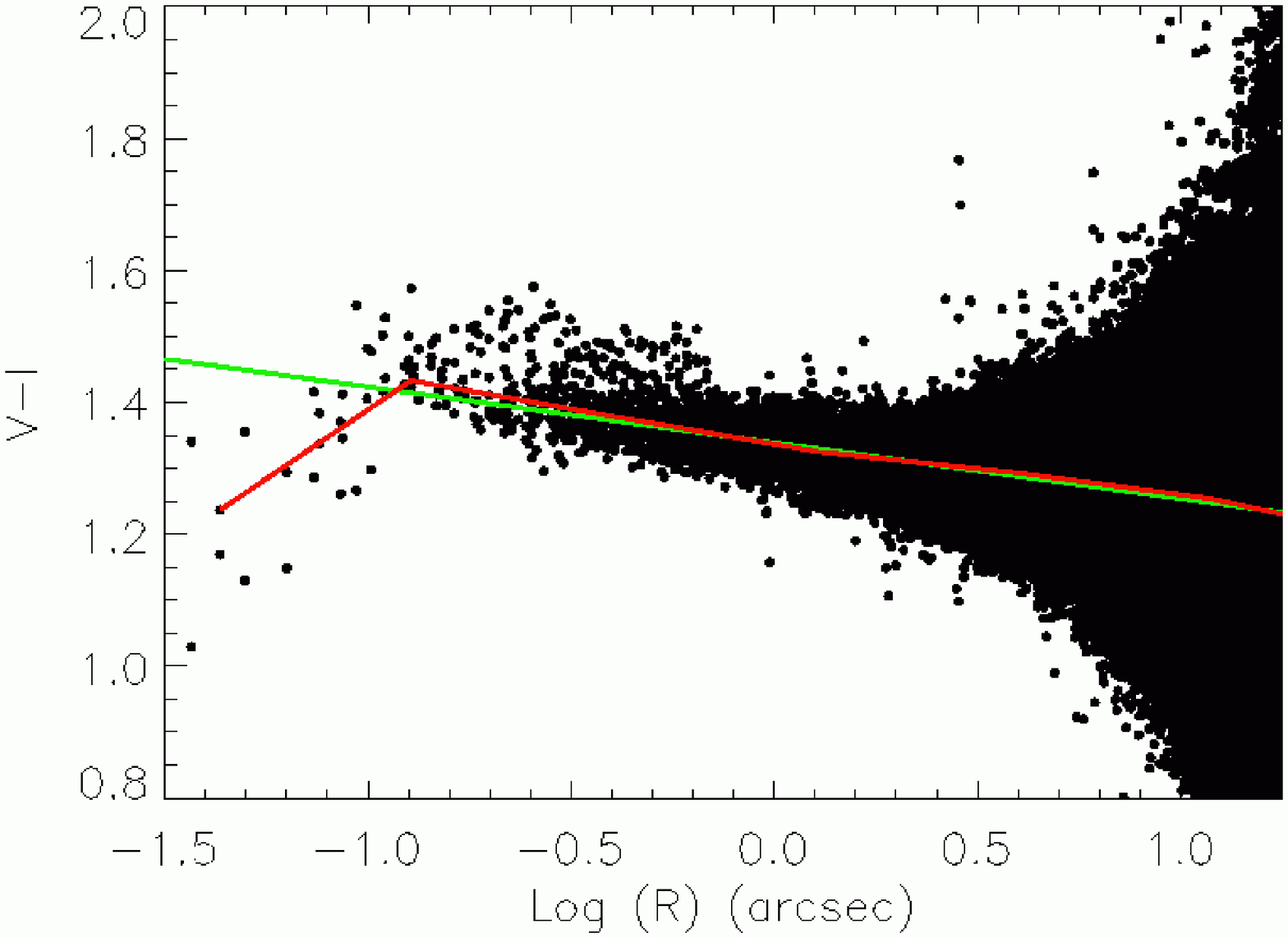}
\caption{The calibrated $V-I$ color of every pixel in the WFPC2/PC1 image of IC~1459 as a function of its elliptical radius. The best-fitting straight line, obtained by minimizing the absolute deviation, is shown in green. For comparison, the red line shows the median color in logarithmically spaced radial bins. Notice the reddened pixels inside $R\la1\arcsec$, caused by dust effects, and the sharp decrease of the color in the nucleus $R\la0\farcs1$, due to a blue source detected previously by \citet{for95a} and C97.\label{fig:ic1459_dust_correction}}
\end{figure}

\begin{figure}
\epsscale{0.5}
\plotone{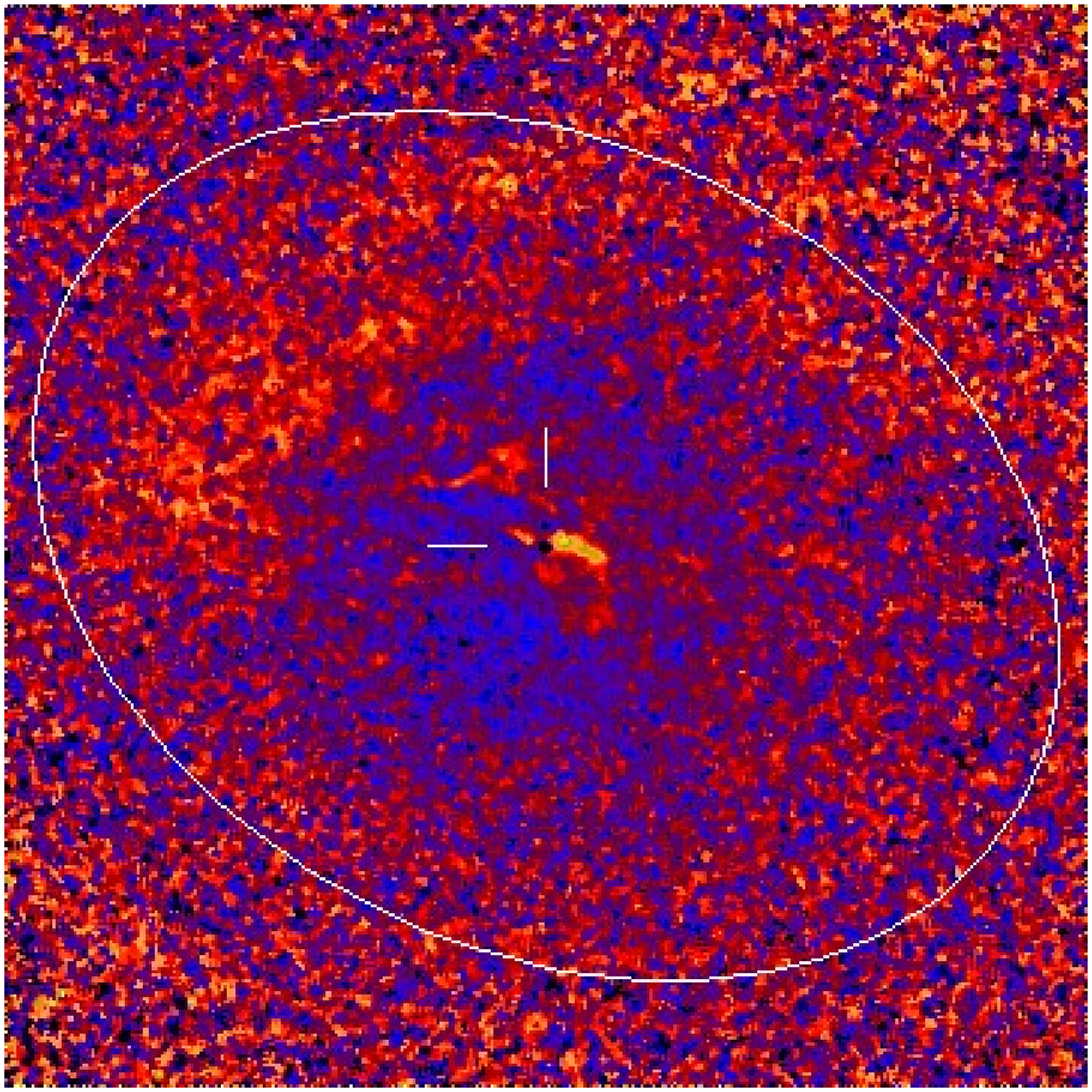}
\caption{$E(V-I)$ map for the central regions of IC~1459, obtained after subtracting the underlying stellar color gradient (see text). Bright colors correspond to large color excesses. Notice the blue central spike (the black dot indicated by the two lines) surrounded by heavy dust absorption (jellow). A representative galaxy isophote, with an 8\arcsec\ semi-major axis, is also shown.\label{fig:ic1459_v-i_image}}
\end{figure}

In practice, C97 corrected the pixel values by assuming that all pixels on the same isophote have the same intrinsic color. We add the assumption that the intrinsic galaxy color varies smoothly as a function of radius. Specifically here we assume the color varies linearly with the logarithm of the radius. This assumption is justified by Figure~\ref{fig:ic1459_dust_correction}. The pixel correction then  involves the following steps:
\begin{enumerate}
\item We measure the average ellipticity and position angle on the original galaxy image.
\item We construct a calibrated color map (in our case $V-I$).
\item We perform a straight line fit to the pixel colors as a function of $\log m$, where $m$ represents the semimajor axis length (see Figure~\ref{fig:ic1459_dust_correction}). The fit is done by minimizing the absolute deviation to reduce the effect of large systematic pixel value deviations due to dust.
\item We compute an $E(V-I)$ map by calculating, for every pixel, the color predicted from the previously fitted relation and subsequently subtracting this from the pixel value. The result is shown in Figure~\ref{fig:ic1459_v-i_image}.
\item We correct the pixels above a given $E(V-I)$ threshold (in general as a function of $m$), using the standard Galactic extinction curve.
\end{enumerate}

This ad hoc method corrects the major effects of patchy dust absorption and can also be used to measure dust corrected color gradients (compare Figure~\ref{fig:ic1459_dust_correction} with Figure~1o in C97).

\section{Dynamical model}
\subsection{The mass model with MGE}

IC~1459 is not a perfectly axisymmetric body, evidenced by a small $\approx5^\circ$ isophotal major axis twist in the range 25\arcsec--100\arcsec\ (\citealt{fra89}; C97). Nevertheless, assuming axisymmetry seems reasonable to study its nuclear dynamics, given that our kinematical observations are all located inside the region that can be well reproduced by assuming axisymmetry.

The first step in the dynamical modeling is to obtain a parametrization of the stellar surface density. For this we adopted the Multi-Gaussian Expansion (MGE) method \citep*{mon92,ems94}, which allows for non-elliptical isophotes and radial ellipticity variations.

We obtained an MGE fit to the WFPC2/F814W dust-corrected images of IC~1459 using the method and software developed by \citet{cap02}. The MGE fit was performed using both the PC1 (at full resolution) and the entire WFPC2 mosaic simultaneously. The PSF was parametrized by fitting a circular MGE model, of the form $PSF(R')=\sum_{k=1}^M G_k \exp{[-R'^2/(2\sigma^{\star 2}_k)]} /(2\pi\sigma^{\star 2}_k)$, to a PSF computed with TinyTim \citep{kri01} for the center of the PC1 CCD. The numerical values of the relative weights $G_k$ (normalized such that $\sum_{k=1}^M G_k=1$), and of the dispersions $\sigma^\star_k$ are given in Table~\ref{tab:mge_psf}.

\begin{figure}
\epsscale{0.65}
\plotone{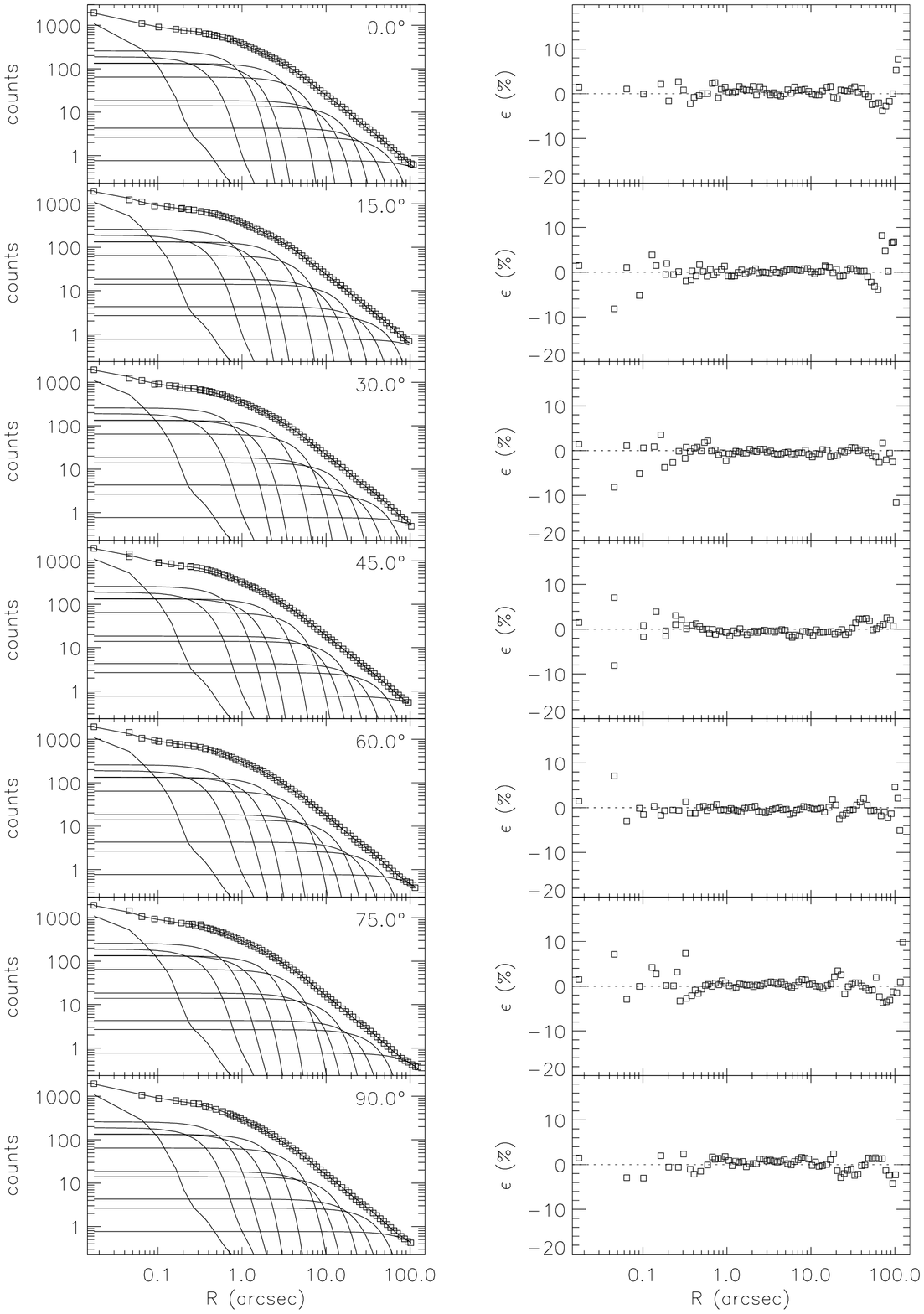}
\caption{\emph{Left Panels:} Comparison between the WFPC2/F814W photometry of IC~1459 in counts (open squares) and the corresponding MGE best fit model (solid line) as a function of radius $R$. The individual convolved Gaussian components are also shown. The fit was performed along 19 sectors, each $5^\circ$ wide, linearly spaced in radius between the major and minor axis. Only every third sector is shown here. \emph{Right Panels:} radial variation of the relative error $\epsilon$ along the profiles. Apart from the very innermost pixels where statistical fluctuations are important, the profiles are reproduced to within 2\%. The RMS error is about 1.0\%.
\label{fig:ic1459_mge_profiles}}
\end{figure}

\begin{figure}
\epsscale{0.5}
\plotone{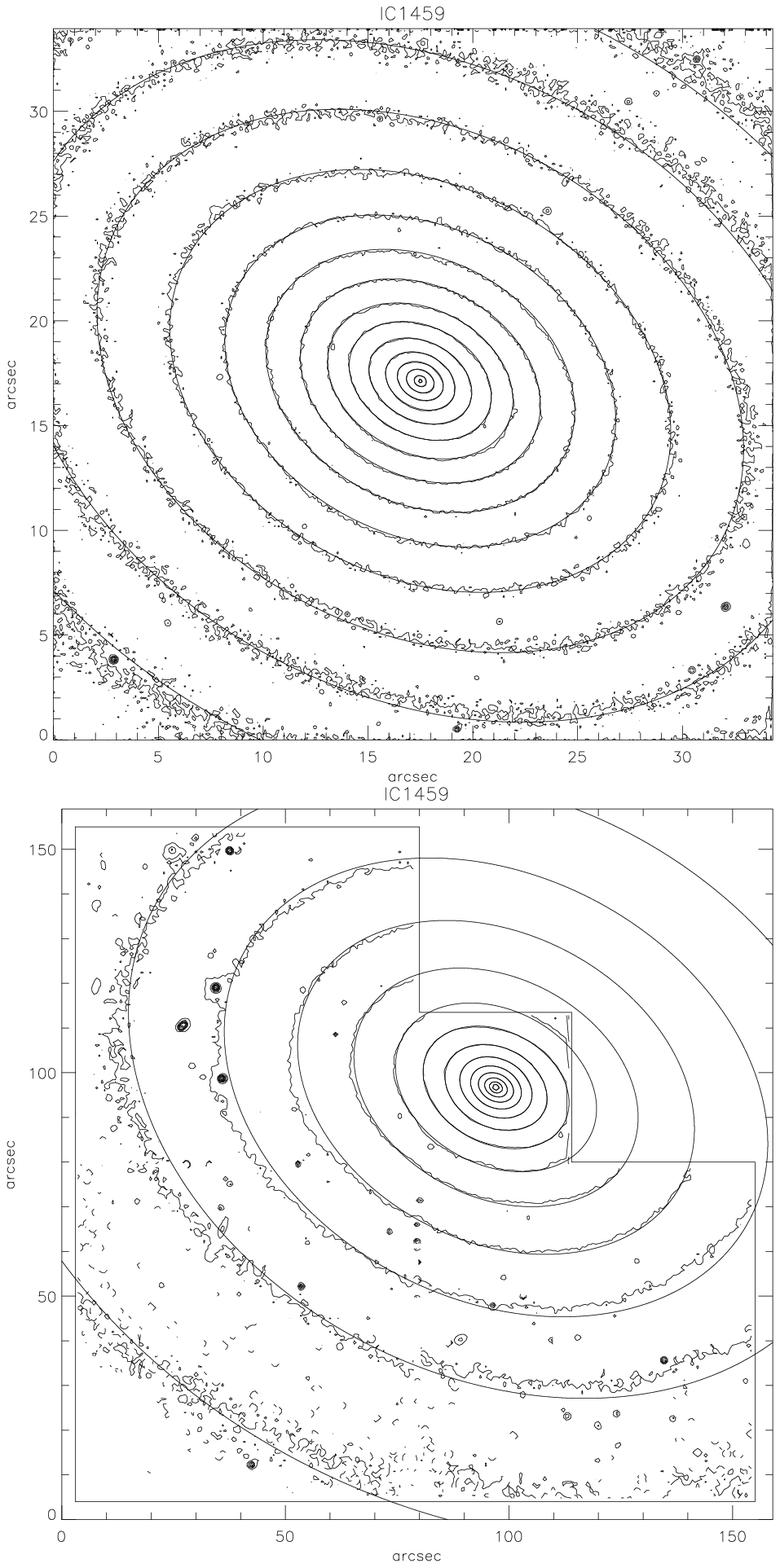}
\caption{Contour maps of the dust-corrected WFPC2 $I$-band image of IC~1459 at two different scales. \emph{Top Panel:} the $35\arcsec\times35\arcsec$ PC1 CCD. \emph{Bottom Panel:} the entire $160\arcsec\times160\arcsec$ WFPC2 mosaic. Superposed on the two plots are the contours of the MGE surface brightness model, convolved with the WFPC2 PSF. The model and data deviate more at larger radii, since the isophotal twist that occurs at these radii cannot be reproduced by an axisymmetric model.
\label{fig:ic1459_mge_contours}}
\end{figure}

Figure~\ref{fig:ic1459_mge_profiles} shows a comparison between the observed photometry and the MGE model along a number of profiles, while Table~\ref{tab:ic1459_mge} gives the corresponding numerical values of the analytically deconvolved MGE parametrization of the galaxy surface brightness
\begin{equation}
\Sigma(x',y')=\sum_{j=1}^N \frac{L_j}{2\pi\sigma_j^2 q'_j}
  \exp\left[-\frac{1}{2\sigma_j^2}\left(x'^2+\frac{y'^2}{q'^2_j}\right)\right].
  \label{eq:mge_surf}
\end{equation}
Also tabulated are the distance-independent central surface brightness of each Gaussian  $I'_j=L_j/(2\pi\sigma_j^2 q'_j)$.
The central Gaussian (no.~1) represents the contribution of the blue unresolved central spike, which is of non-thermal origin and will not be included in the dynamical model calculation. The comparison between the isophotes of the convolved model and the actual image is shown in Figure~\ref{fig:ic1459_mge_contours}. We confirm the finding of C97 that there is no photometric evidence for a nuclear disk in the dust corrected images.

\begin{table}
\caption{Parameters of the HST/WFPC2/F814W circular MGE PSF
\label{tab:mge_psf}}
\centering
\begin{tabular}{ccc}
\tableline\tableline
$k$   &   $G_k$ & $\sigma_k^\star$ \\
      &          & (arcsec)  \\
\tableline
1  &  0.294  &   0.0224  \\
2  &  0.559   &   0.0655  \\
3  &  0.0813   &   0.214   \\
4  &  0.0657   &   0.610   \\
\tableline
\end{tabular}
\end{table}

\begin{table}
\caption{MGE parameters for the deconvolved \emph{I}-band
WFPC2 surface brightness of IC~1459
\label{tab:ic1459_mge}}
\centering
\begin{tabular}{ccccc}
\tableline\tableline
$j$   &   $I'_j$ & $\sigma_j$ & $q'_j$ & $L_j$ \\
      &   (\lsun$_{,I}$ pc$^{-2}$) & (arcsec) &  & ($10^9$ \lsun$_{,I}$) \\
\tableline
1  &  727874  &   0.0172  &   1.000 & 0.0295 \\
2  &  18191   &   0.268   &   0.899 & 0.159 \\
3  &  22306   &   0.618   &   0.672 & 0.777 \\
4  &  11511   &   0.993   &   0.815 & 1.25 \\
5  &  11226   &   2.09    &   0.681 & 4.55 \\
6  &  5243.0  &   3.77    &   0.775 & 7.82 \\
7  &  1577.4  &   6.82    &   0.714 & 7.10 \\
8  &  1105.6  &   11.2    &   0.738 & 13.9 \\
9  &  337.28  &   21.8    &   0.732 & 15.9 \\
10 &  194.31  &   39.6    &   0.733 & 30.3 \\
11 &  59.276  &   126    &    0.774 & 99.5 \\
\tableline
\end{tabular}
\end{table}

\subsection{Three-integral models}

We model IC~1459 by using an axisymmetric three-integral implementation of Schwarz\-schild's (1979) orbit superposition method, developed by \citet{rix97,van98,cre99b} and \citet{ric02}. The method is described extensively  in the above references and we will not repeat the details here. Very briefly, the stellar density is derived by deprojecting (see Appendix~\ref{sec:potential}) an MGE parametrization of the observed surface density (equation~[\ref{eq:mge_surf}]), assuming axisymmetry and a value for the stellar mass-to-light ratio $\Upsilon$. Stellar orbits are integrated in the potential that is generated by both the stellar density and a central supermassive BH. Each orbit is projected onto the space of the observables, taking into account PSF convolution and aperture binning. Finally, the distribution of orbital weights that best fits both the luminosity density distribution of the model and the observed kinematics is determined by solving a Non-Negative Least-Squares (NNLS, \citealt{law74}) problem.

We modified the software implementation by \citet[hereafter M98]{van98} such that it can deal with the MGE parametrization of the surface brightness derived in the previous paragraph, in a way that is similar to \citet{cre99a}. The differences between our software and the implementation used by M98 are:
\begin{enumerate}
\item Simplified calculation of the potential. This is possible because of the MGE formalism  and the details described in Appendix~\ref{sec:potential};
\item Substitution of the numerical integration of the luminosity density along the line of sight with the analytic MGE projection;
\item Simplified and more accurate (due to the smaller number of numerical integrations involved) evaluation of the density integration on a polar or Cartesian grid (Appendix~\ref{sec:integration}).
\end{enumerate}

\subsection{Orbit library}

To include a representative set of orbits in the orbit library, we sample them on a grid that covers the full extent of the integral space $(E,L_z,I_3)$, where $E$ is the energy, $L_z$ is the angular momentum along the $z$ symmetry axis and $I_3$ is a non-classical integral. This is achieved by sampling the energy logarithmically in the radius of the circular orbit with the given energy $E(R_c)$, from 0\farcs05 to 300\arcsec. This grid includes 99\% of the total luminous mass of the galaxy. The quantity $\eta\equiv L_z/L_{\rm max}$ is sampled linearly on an open grid in the interval $[0,1]$. $I_3$ is sampled linearly in the angle $w$ on an open grid in the interval $[0,w_{\rm th}]$, where $w_{\rm th}$ is the angle $w$ for the ``thin tube'' orbit at the given $(E,L_z)$ (see Figure~5 in M98 for details).  After some testing we chose a $20\times14\times7$ grid in integral space (including both positive and negative $L_z$), but similar results (with longer computation times) were obtained with a $25\times26\times13$ grid. This choice of parameters is similar to that used in the modeling of other nearby galaxies (e.g. van der Marel et al. 1998; Verolme et al. 2002).

\section{Results}
\subsection{Stellar kinematics}
\label{sec:stell_kin}

Our model has three free parameters: the galaxy inclination $i$, the BH mass $M_\bullet$ and the stellar mass-to-light ratio $\Upsilon$. Inclinations smaller than $i\la55^\circ$ can be excluded based on the observed axial ratio of IC~1459, since these values correspond to intrinsic stellar densities that are too flat for an elliptical galaxy.

\begin{figure}
\epsscale{0.4}
\plotone{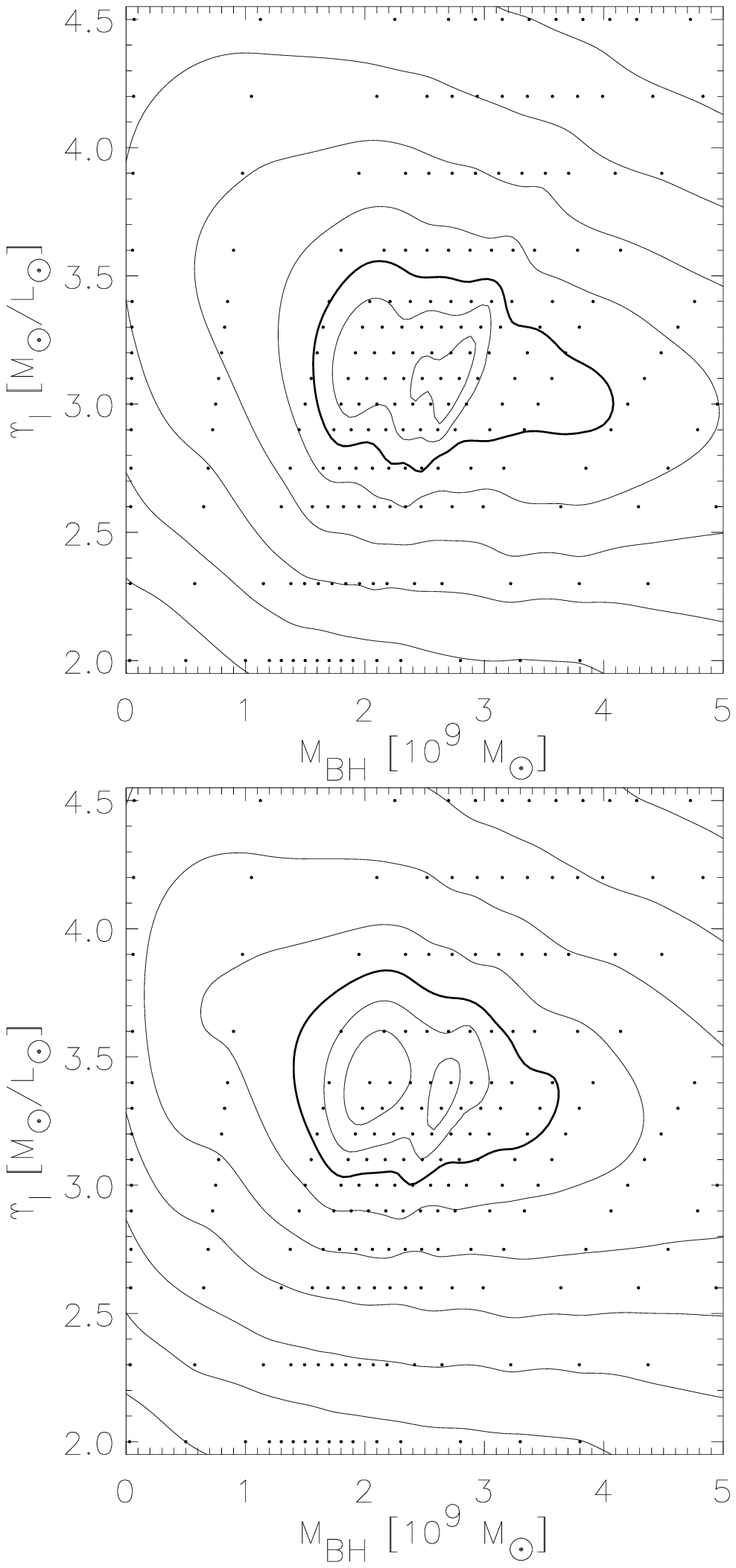}
\caption{\emph{Top Panel:} Contours of constant $\chi^2$, measuring the goodness of fit of the edge-on a\-xi\-symmetric models for IC~1459. The abscissa is the mass $M_\bullet$ of a point mass representing a nuclear BH; the ordinate is the constant stellar mass-to-light ratio $\Upsilon$. These two parameters uniquely determine the potential given the inclination $i$. Every dot corresponds to a three-integral axisymmetric dynamical model. The contours were obtained through a minimum curvature interpolation. The first three contours define the formal 68.3\%, 95.4\% and (heavy contours) 99.73\% confidence regions for the three parameters $(i,M_\bullet,\Upsilon)$ jointly; subsequent contours are characterized by a factor of 2 increase in $\Delta\chi^2$. The best fit is obtained with a mass for the BH $M_\bullet\simeq2.6\times10^9 M_\odot$ and a stellar $\Upsilon\simeq3.1$ in the $I$-band. \emph{Bottom Panel:} Same as in top panel for a model with  $\Upsilon$ varying as a function of radius (see text for details). Here the ordinate represents the value of  $\Upsilon$ at a radius $R=1\arcsec$.
\label{fig:chi2}}
\end{figure}

\begin{figure}
\epsscale{1}
\plotone{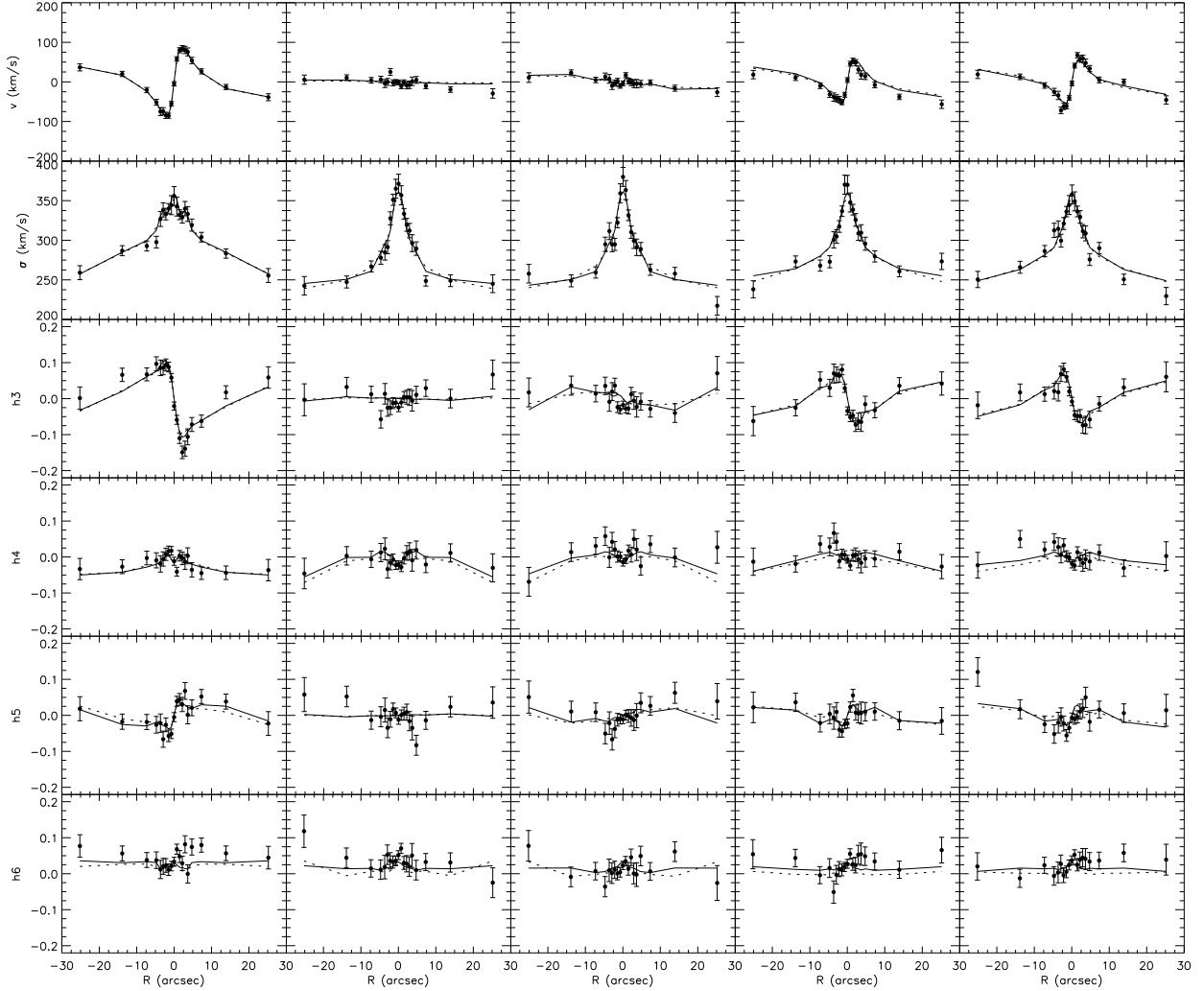}
\caption{Data-model comparison. From top to bottom: the mean velocity $V$, the velocity dispersion $\sigma$ and the higher order Gauss-Hermite moments ($h_3-h_6$) as a function of position along the slit. From left to right: the measurements along the major (PA=39$^\circ$), minor (PA=128$^\circ$, 120$^\circ$) and intermediate axis (PA=83$^\circ$, 173$^\circ$). The filled circles with error bars represent the measurements, while the solid line gives the prediction of the best fitting unregularized dynamical model, taking into account seeing and pixel binning. The dashed line shows the prediction of the high-regularization solution ($\Delta=1$). The case of small regularization  ($\Delta=4$) lies between the two curves.
\label{fig:plot_fit_major}}
\end{figure}

\begin{figure}
\epsscale{0.5}
\plotone{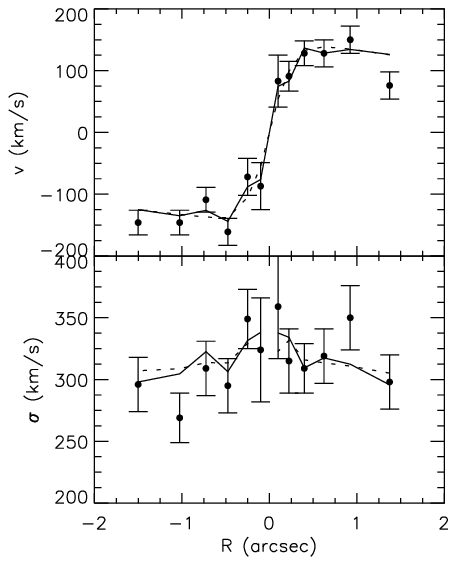}
\caption{The fit to the STIS/G430L kinematic observations for the same model that is shown in Figure~\ref{fig:plot_fit_major}. The upper panel displays the mean velocity $V$, while the bottom panel is the velocity dispersion $\sigma$, both as a function of radius along the major axis. The filled circles with error bars represent the measurements, while the solid line gives the prediction of the best fit unregularized dynamical model. The dashed line shows the predictions of the high regularization solution ($\Delta=1$). As in Figure~\ref{fig:plot_fit_major}, the case of small regularization  ($\Delta=4$) lies between the two curves.
\label{fig:plot_fit_stis}}
\end{figure}

We fitted dynamical models to the $V$, $\sigma$ and $h_3$--$h_6$ observations along all five position angles and to the STIS/G430L $V$ and $\sigma$ measurements. We varied the parameters $\Upsilon$ and $M_\bullet$ for three different values of the inclination $i=60^\circ$, $i=75^\circ$ and $i=90^\circ$. The overall best fit was found for an inclination $i=90^\circ$ (corresponding to edge-on viewing). Contours of equal $\chi^2$, using the confidence levels for a $\chi^2$-distribution with three degrees of freedom, are shown in the upper panel of Figure~\ref{fig:chi2}. The best-fit parameters are a mass-to-light ratio of $\Upsilon=3.1\pm0.4$ (in the $I$-band) and a central black hole mass of  $M_\bullet=(2.6\pm1.1)\times10^9 M_\odot$\ (3$\sigma$ level). The inclination is not strongly constrained by our data and our results do not depend significantly on the adopted value. In particular, the best-fitting black hole mass at $i=60^\circ$ is $M_\bullet=(2.5\pm1.2)\times10^9 M_\odot$. The best-fit model has $\chi^2\approx867$ with $N=594$ kinematical constraints. The fact that $\chi^2>N$ even for the best model, is due to some small systematic errors, which are in particular visible for the even Gauss-Hermite moments $h_4$ and $h_6$. The excellent match (Figure~\ref{fig:plot_fit_major}, \ref{fig:plot_fit_stis}) to the observed kinematical constraints along all position angles suggests that the decoupled core in IC~1459 can be well described as an axisymmetric component. This allows us to study its internal dynamics, shape and mass distribution.

The solid lines in Figure~\ref{fig:plot_fit_major} and \ref{fig:plot_fit_stis} were calculated by fitting to only the photometry and kinematics. The kinematical constraints are distributed over 95 apertures (not all independent), each of which provides 6 Gauss-Hermite moments and 12 STIS apertures with $V$ and $\sigma$ measurements. Selfconsistency is ensured by fitting to the mass in the meridional (120 constraints) and projected plane (140 constraints), as well as to the mass fractions contained in the kinematical apertures. The models therefore have to satisfy a total of 961 constraints, which are to some extent correlated. Since our library contains $20\times14\times7=1960$ orbits whose orbital weights are the unknowns that have to be determined, there is no unique solution that defines the internal dynamics of this galaxy (in the approximation of this model). In fact, the NNLS matrix equation that is solved is numerically rather ill-conditioned, giving rise to a distribution function composed of sharp isolated peaks. Such DFs are unphysical. A smoother and better (see Cretton et al. 1999; Verolme \& de Zeeuw 2002) solution can be obtained by using regularization. The regularization scheme that is used here minimizes, up to a certain degree, the differences in weights between neighboring orbits. Although these additional regularization constraints result in a slightly poorer fit to the data, a considerable amount of smoothing can be forced upon the solution before systematic deviations in the fit become apparent.

\begin{figure}
\epsscale{1}
\plotone{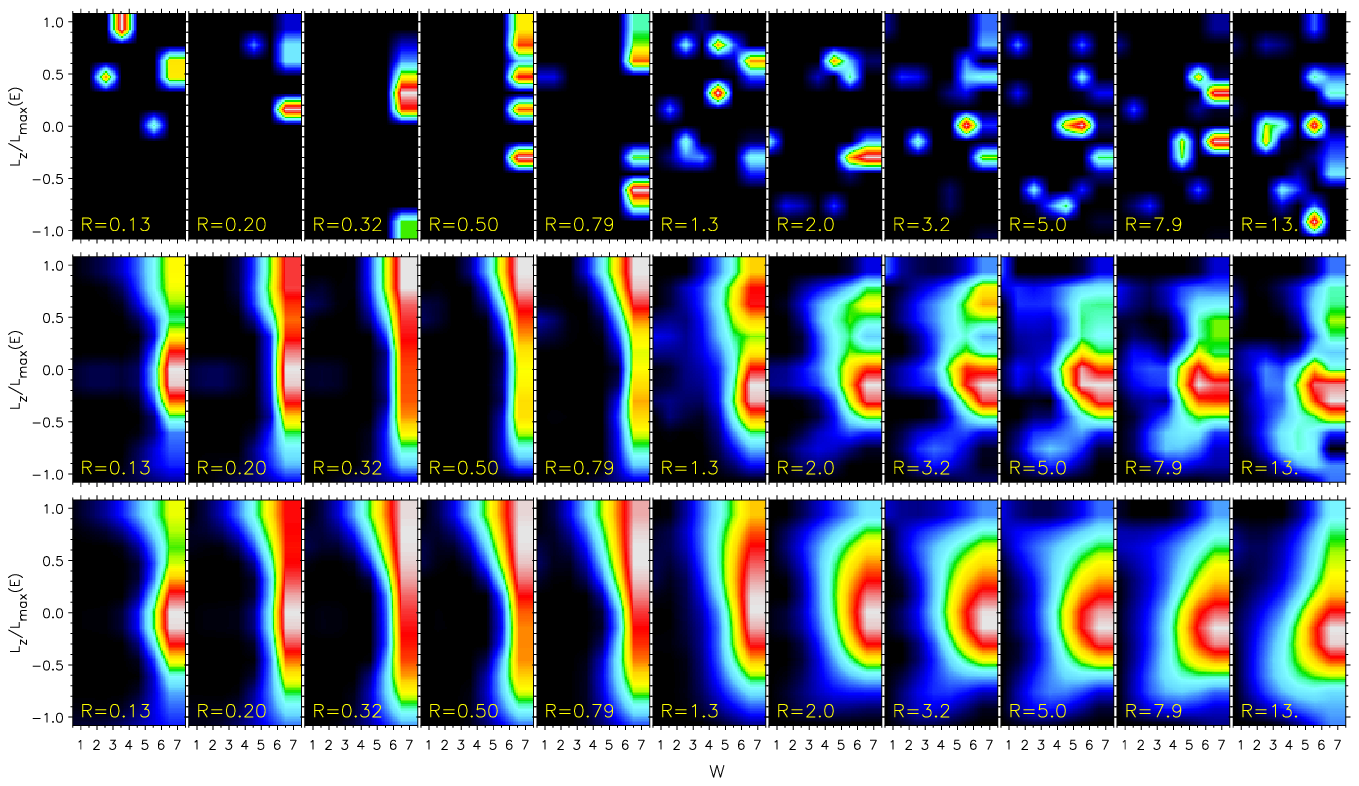}
\caption{The top row shows the $(\eta, w)$ integral space (defined as described in the text) at a selected set of energies, for the best-fitting non-smooth orbit-superposition model. The radius of the circular orbit (in arcsec) at any given energy is printed at the bottom of each panel and only the radii that are actually constrained by the observed data are shown. The plots were obtained by bilinear interpolation on a $14\times7$ grid in $(\eta,w)$, where each $(\eta,w)$ represents an orbit. Given that the value of $w$ does not have direct physical significance we labeled the abscissa with the bin number from 1--7. The colors show the fractional mass that was assigned to the orbit by the NNLS fit (bright colors correspond to large weights). The weights are normalized to the maximum value at the corresponding energy. Smoother solutions are obtained by adding regularization constraints to the NNLS fit. The second and third rows show the integral space for the same model with a small ($\Delta=4$) and a large ($\Delta=1$) amount of regularization, respectively.  Negative values of $\eta\equiv L_z/L_{\rm max}(E)$ correspond to the bulk rotation of the galaxy body.  The counterrotating component is clearly recognizable as a well-separated peak at $\eta\approx1$ in the central and bottom row, in particular in the radial range 0\farcs3--0\farcs8 (panels 3--5 from the left).
\label{fig:ic1459_integral_space}}
\end{figure}

Figure~\ref{fig:ic1459_integral_space} shows the integral space [the weights on the $(E,L_z,I_3)$ grid] of our best-fitting model, for a selected set of energies (corresponding to the radii effectively constrained by the observations). The top row shows the integral space for the unsmoothed solution which, as a consequence, appears very noisy. A small amount of regularization ($\Delta=4$; see M98 for details) was added as a constraint in the middle row. While the fit to the data remains essentially unchanged, some coherent structures clearly emerge in the solution space. The bulk of the galaxy rotation can be recognized as a large bright (high weight) region at low and negative $\eta\equiv L_z/L_{\rm max}$ and high $w$ on the range $\approx$1\arcsec--13\arcsec. At radii smaller than $\approx$0\farcs8, the regularized solution contains a well-separated group of nearly circular orbits ($\eta\approx1$), which counterrotates with respect to the main galaxy body.
These orbits appear necessary to explain the observed rapid counterrotation in the kinematics. The contribution of the counterrotating component seems to decrease inside $R\lesssim0\farcs2$. This is probably not a real effect: there is not enough data inside this radius to constrain the phase space in detail and its appearance is dictated by the smoothness constraint.

The structure of integral space remains qualitatively the same at high regularization ($\Delta=1$, bottom row). This latter model is still able to produce a reasonable fit to all constraints (see dashed line in Figure~\ref{fig:plot_fit_major}, \ref{fig:plot_fit_stis}).
Although our conclusions do not depend strongly on the regularization parameter $\Delta$, in the following we adopt $\Delta=4$ for our plots. Both Cretton et al. (1999) and Verolme et al. (2002) have carried out a number of tests of the Schwazrschild code, and established which value of $\Delta$ gives the best reconstruction of a known distribution function. The result is a $\Delta\sim4$. Detailed comparison with independent results obtained by Gebhardt with a maximum entropy technique (which starts at the opposite end with a fully smooth distribution function, and then makes it less and less so) shows that again an amount of smoothing equivalent to our $\Delta\sim4$ is optimal.

Both the best fit values and the details of the internal kinematic structure are in excellent agreement with our previous results \citep{cap00,cap01}, which started from a triple power-law parametrization of the stellar surface density instead of the MGE expansion and were constrained by ground-based data only. This comparison shows that the best-fitting parameters do not depend on the details of the adopted stellar density parametrization. However, our best-fitting black hole mass is slightly larger than the value $(1.1\pm0.3)\times10^9 M_\odot$\ that is predicted by the $M_\bullet$-$\sigma$ relation \citep{fer00,geb00} as given by Tremaine et al. (2002).

The contribution of a dark halo becomes evident in the observed stellar kinematics only well beyond an effective radius, and hence can be safely ignored (Carollo et al.\ 1995; Gerhard et al.\ 2001).
Furthermore, we follow standard practice and assume that the stellar mass-to-light ratio does not vary with radius.
To determine whether our results depend on this assumption, we also ran models in which $\Upsilon$ follows the change of the $V-I$ color presented in Fig.~\ref{fig:ic1459_dust_correction}. The color gradient of IC~1459 is $d(V-I)/d(\log R)\approx-0.084$. Using models from Charlot, Worthey \& Bressan (1996), the color variation can be translated into a fractional decrease of $\Upsilon$ by $\sim15$ per cent per decade in radius, more or less independent of whether the color gradient  is due to age or to metallicity variations. The results obtained from models with a non-constant $\Upsilon$ do not show any significant difference in the best fitting BH mass (bottom panel of Figure~ \ref{fig:chi2}) or in the internal dynamical structure, compared to the models with constant $\Upsilon$.

\begin{figure}
\epsscale{0.6}
\plotone{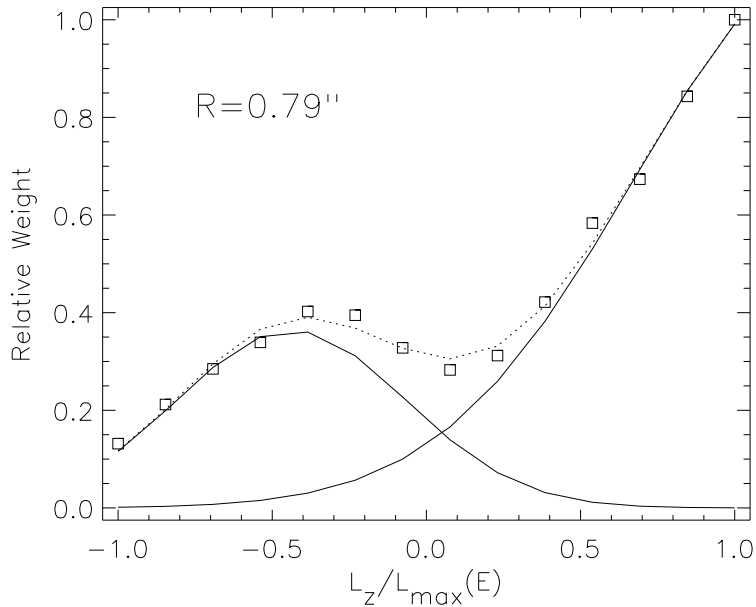}
\caption{The open squares show the mass fraction as a function of angular momentum, for orbits with an energy corresponding to a circular orbit of radius $R_c=0\farcs79$. A small amount of regularization ($\Delta=4$) was used. The solid lines represent a double Gaussian fit to the mass fractions, while the dashed line is the corresponding sum. This plot illustrates the fact that we can
make a distinction between the phase-space contribution of the counterrotating component, which peaks at $L_z/L_\textrm{max}\sim1$, and the bulk of the galaxy stars (centered around $L_z/L_\textrm{max}\sim-0.4$).
\label{fig:two_components}}
\end{figure}

Given that the counterrotating component appears so well isolated from the rest of the orbital distribution (see also Figure~\ref{fig:two_components}), we can make a quantitative estimate of the mass involved. This is not possible by any other means, due to the fact that the counterrotating component is not apparent from the photometry and does not appear as a double component in the LOSVD.  By summing the solution weights of the orbits that contribute to the counterrotation, we find that this component carries $\approx0.5$\% of the luminous galaxy mass which is $\approx3\times10^9 M_\odot$  (determined from the Gaussian components of Table~\ref{tab:ic1459_mge}, using the best fitting mass-to-light ratio).

\begin{figure}
\epsscale{0.6}
\plotone{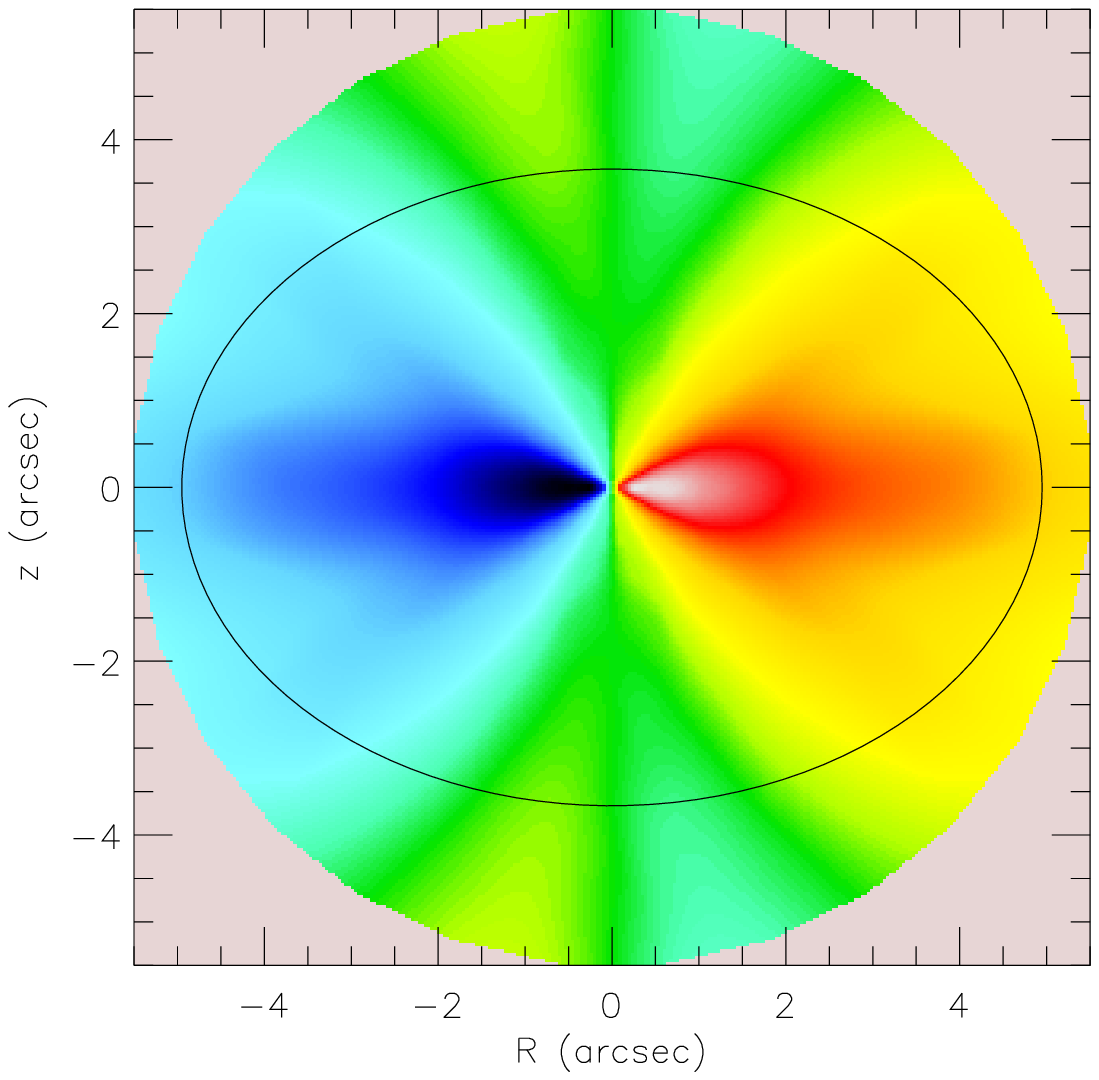}
\caption{The $V/\sigma$ field in the meridional plane, computed from the velocity moments of the best fitting low regularization ($\Delta=4$) solution of Figure~\ref{fig:ic1459_integral_space}. White and black correspond to $V/\sigma=\pm0.3$, while green indicates the zero velocity curve. The flattened appearance of the $R\la2$\arcsec\ counterrotating component is readily apparent. A representative galaxy isodensity contour is also shown. This field was computed from the orbital solution moments on a complete grid in the galaxy meridional plane and was \emph{not} interpolated from the values in a few aperture positions (as done in Figure~\ref{fig:ic1459_kinematics}).
\label{fig:ic1459_merid_velfield}}
\end{figure}

\begin{figure}
\epsscale{0.5}
\plotone{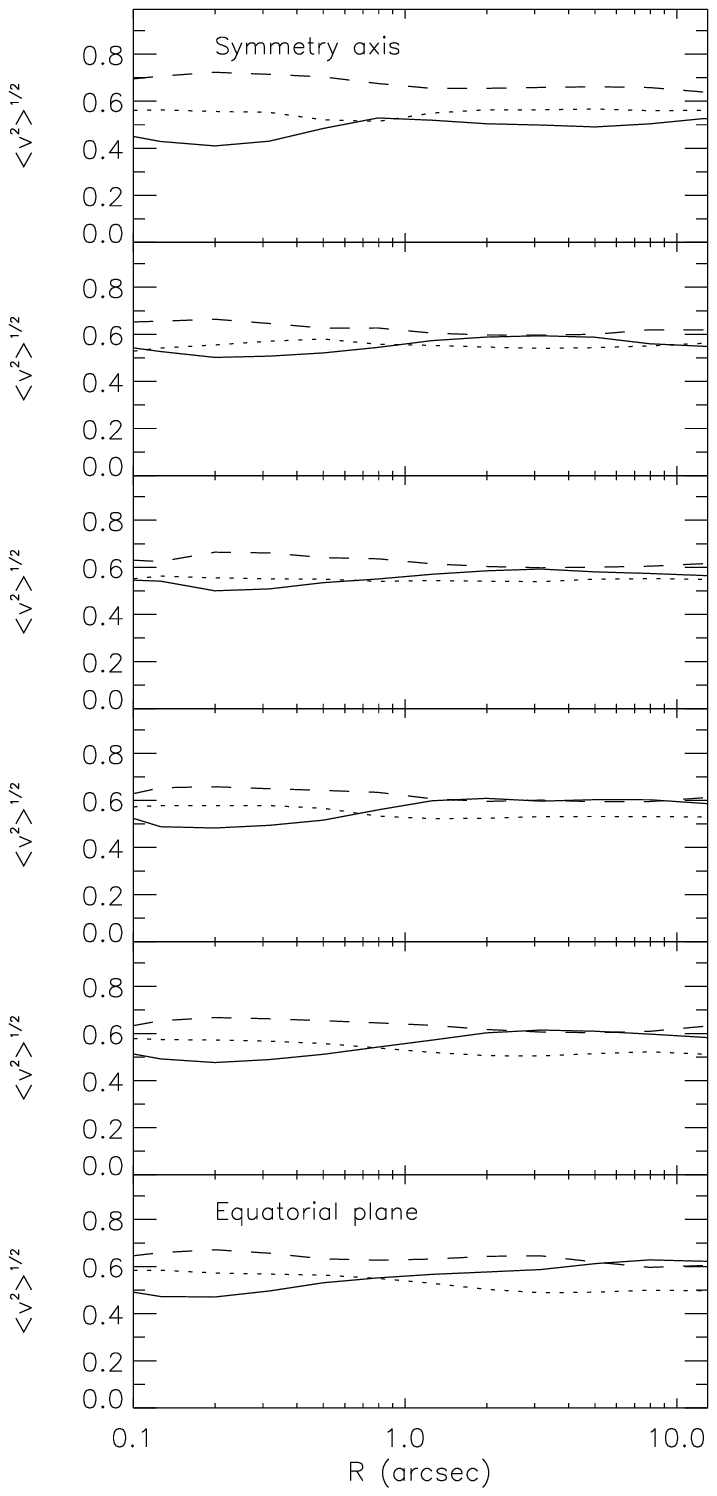}
\caption{The intrinsic second velocity moments $\left<v_r^2\right>^\frac{1}{2}$ (solid line), $\left<v_\theta^2\right>^\frac{1}{2}$ (dotted line) and $\left<v_\phi^2\right>^\frac{1}{2}$ (dashed line) are plotted as a function of the polar radius in the meridional plane, along sectors linearly spaced in angle between the symmetry axis (upper panel) and the equatorial plane (bottom panel). The moments are normalized by the total RMS velocity $\left<v^2\right>^\frac{1}{2}$.
\label{fig:ic1459_moments_ratios}}
\end{figure}

Figure~\ref{fig:ic1459_merid_velfield} shows the $V/\sigma$ field in the meridional plane of the model. This was calculated by adding the velocity moments of the orbits in the best fitting solution with low regularization of Figure~\ref{fig:ic1459_integral_space}. The flattened appearance of the counterrotating stellar component is clearly visible. Figure~\ref{fig:ic1459_moments_ratios} presents the velocity moments in polar coordinates. We see the nuclear kinematics shows mild tangential anisotropy, which compares well with the corresponding measurements obtained for other galaxies (Gebhardt et al.\ 2002).
However, the underlying structure is more complex as the distribution function is in fact the sum of the bulk of the galaxy and the counter rotating component, which is not evident in the second moments.

\citet{cre99a} showed that the Gauss-Hermite (GH) parame\-trization for the LOSVD, which is used to constrain the kinematics, can produce artificial counterrotations in some special cases. This is essentially due to the fact that the model is trying to reproduce a small number of GH moments and not the actual observed LOSVD and when two functions coincide on a small set of GH moments this does not guarantee that they are the same function. This phenomenon can be easily recognized after the fit, however, by comparing the LOSVDs predicted by the model and the observed ones. Before interpreting the internal dynamics of IC~1459, we check that our best fit dynamical model does not suffer from this shortcoming.

\begin{figure}
\epsscale{0.6}
\plotone{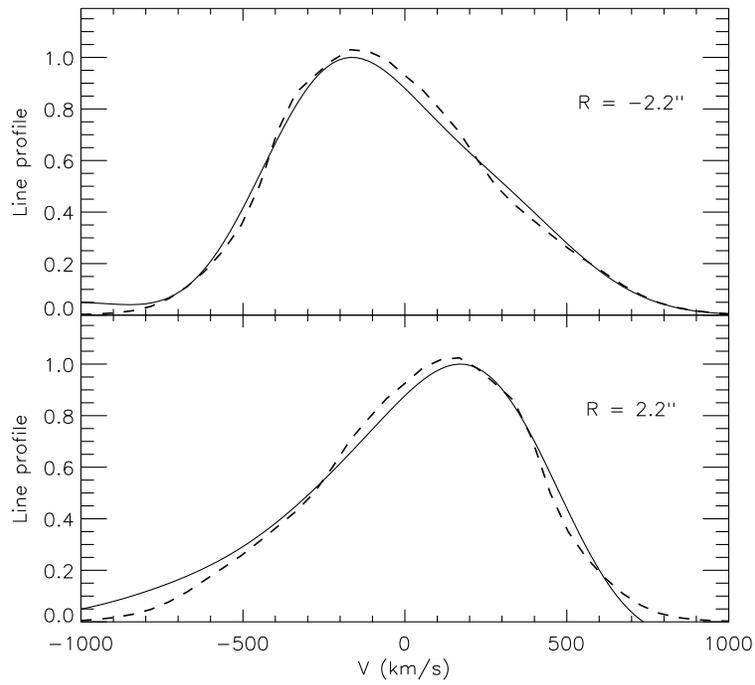}
\caption{The LOSVD predicted by the models with low regularization ($\Delta=4$, dashed line) at the radius of maximum observed rotation ($R=\pm2\farcs2$), together with the observed LOSVD, which is reconstructed from the first six Gauss-Hermite moments (solid line).
\label{fig:ic1459_losvd}}
\end{figure}

Figure~\ref{fig:ic1459_losvd} shows the LOSVD of the model with low regularization ($\Delta=4$), obtained by direct summation on the velocity histograms of the best fitting orbits, together with the observed LOSVD. The latter was reconstructed from the Gauss-Hermite parametrization introduced by \citet{van93}, using the mean velocity, the velocity dispersion and the Gauss-Hermite moments that are plotted in Figure~\ref{fig:plot_fit_major} and \ref{fig:plot_fit_stis}. The Hermite polynomials are tabulated in the Appendix~C. The plot only represents the most extreme observed LOSVD, but similar results are obtained at the other ground-based apertures. It is apparent from this comparison that the model provides an excellent fit to the observed LOSVD, at least well within the errors suggested by the observed differences in the LOSVD at opposite sides of the galaxy, that are visible in Figure~\ref{fig:ic1459_losvd}. We can thus safely affirm that we are \emph{not} in the situation of Figure~10 of \citet{cre99a} and that our model reproduces the data correctly.

\subsection{The STIS gas kinematics}

The very central STIS/G430L spectrum of IC~1459 covers the wavelength
range between 3500 and 5800 \AA. It is dominated by an unresolved
source of featureless, non-stellar continuum and by prominent and
broad ($\sigma\sim500$ \kms) forbidden gas emission lines of
[O~{\scriptsize II}]~$\lambda$3727, [Ne~{\scriptsize
  III}]~$\lambda$3869, [S~{\scriptsize II}]~$\lambda\lambda$4068,4076,
[O~{\scriptsize III}]~$\lambda$4363, [O~{\scriptsize
  III}]~$\lambda\lambda$4959,5007, [N~{\scriptsize I}]~$\lambda$5200
and the Balmer series emission-lines H$\delta$, H$\gamma$ and
H$\beta$. A detailed analysis of the physical properties of this
spectrum will be presented elsewhere. In this Section we focus on the
gas kinematics.

\begin{figure}
\epsscale{0.55}
\plotone{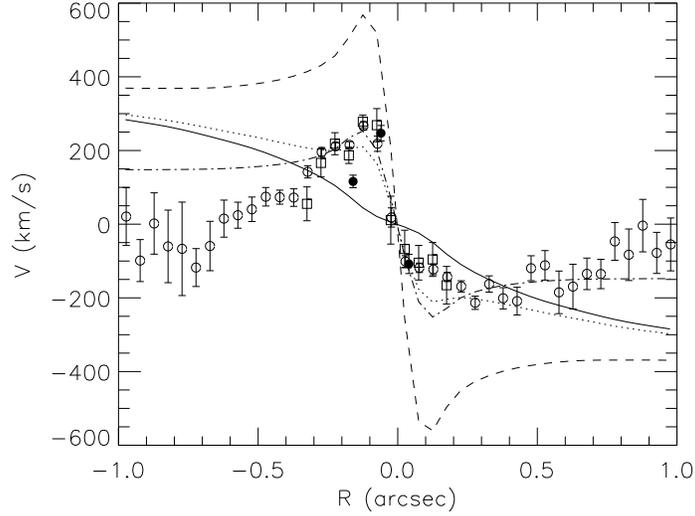}
\caption{Models of the STIS/G430L gas kinematics. The open circles with error bars represent the observed mean velocity of the gas as measured from the [O~{\scriptsize
II}]~$\lambda$3727 line, the open squares are the measurements from the H$\beta$ line. Also shown with the filled circles are the measurements by VK00, derived from the H$\alpha$+[N~{\scriptsize II}] blend on the FOS spectra, through the 0\farcs086 square apertures. The solid, dotted and dashed lines are the predicted model velocities assuming a disk inclination of $i=60^\circ$, $\theta=0^\circ$ and a
central black hole masses of 0, $3.5\times10^8 M_\odot$ and $2.6\times10^9 M_\odot$,  respectively. The dash-dotted line is the predicted velocity for a model with $i=20^\circ$, $\theta=0^\circ$ and $M_\bullet=2.6\times10^9 M_\odot$. \label{fig:gas_modeling}}
\end{figure}

\begin{figure}
\epsscale{0.55}
\plotone{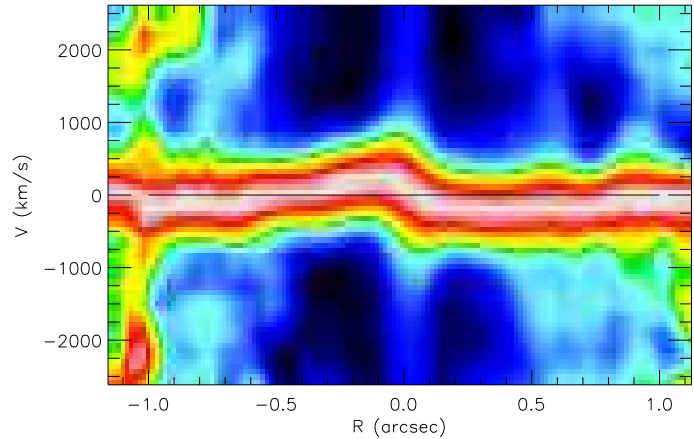}
\caption{The observed [O~{\scriptsize II}]~$\lambda$3727 line on the STIS/G430L spectrum. To make the gas velocity curve more visible, each column of the spectrum was divided by the maximum value of the emission line in that column. The spectrum was then resampled on a $3\times$ finer grid by means of bilinear interpolation. \label{fig:oii_line}}
\end{figure}

Of all emission lines, [O~{\scriptsize II}]~$\lambda$3727 can be
traced reliably to the largest distance from the center
($\sim$1\arcsec). Single Gaussians provide a good fit to the emission
lines and are therefore used to determine the mean velocity and
velocity dispersion at each row of the spectrum
(Figure~\ref{fig:gas_modeling}, Table~\ref{tab:gas_kin}). The reality of the velocity curve that was
determined in this manner can be verified in Figure~\ref{fig:oii_line}. The
line intensity was normalized by dividing each column of the spectrum
by the maximum intensity of the [O~{\scriptsize II}] line in that
column. The mean velocity curves for the other species agree within
the errors with the [O~{\scriptsize II}] line in the regions of
overlap. The galaxy nucleus is assumed to be at the peak in the
continuum flux. The most salient features are a central steep velocity
gradient in the direction opposite to the nuclear stellar rotation and
a clear asymmetry of the velocity curve with respect to the nucleus.
The velocity curve reaches a peak velocity difference of $\pm200$
\kms, then drops rapidly to zero and even changes sign at
$\sim0\farcs5$ on one side of the nucleus.

\clearpage
\begin{deluxetable}{rrrrr}
\tablecolumns{5}
\tablewidth{0pc}
\tablecaption{IC~1459, PA=34$^\circ$, STIS gas kinematics.\label{tab:gas_kin}}
\tablehead{
\colhead{R} & \colhead{$V$} & \colhead{$\Delta V$} & \colhead{$\sigma$} & \colhead{$\Delta \sigma$}  \\
\colhead{(\arcsec)} & \colhead{(\kms)}  & \colhead{(\kms)} & \colhead{(\kms)} & \colhead{(\kms)}
}
\startdata
\cutinhead{[\ion{O}{2}] $\lambda$3727 emission line}
  -0.97 &     20 &      78 &      327 &       79 \\
  -0.92 &    -98 &      57 &      243 &       57 \\
  -0.87 &      1 &      83 &      159 &       83 \\
  -0.82 &    -60 &      98 &      202 &       99 \\
  -0.77 &    -66 &     126 &      218 &      128 \\
  -0.72 &   -117 &      51 &      139 &       51 \\
  -0.67 &    -59 &      66 &      307 &       67 \\
  -0.62 &     15 &      50 &      127 &       50 \\
  -0.57 &     24 &      35 &      172 &       35 \\
  -0.52 &     40 &      33 &      179 &       33 \\
  -0.47 &     74 &      25 &      148 &       25 \\
  -0.42 &     73 &      19 &      207 &       19 \\
  -0.37 &     72 &      23 &      284 &       24 \\
  -0.32 &    142 &      16 &      271 &       16 \\
  -0.27 &    195 &      12 &      283 &       12 \\
  -0.22 &    211 &       9 &      295 &        9 \\
  -0.17 &    215 &       8 &      302 &        9 \\
  -0.12 &    266 &      11 &      344 &       11 \\
  -0.07 &    218 &      20 &      423 &       20 \\
  -0.02 &     18 &      26 &      470 &       26 \\
   0.02 &   -101 &      15 &      411 &       15 \\
   0.07 &   -118 &      12 &      326 &       13 \\
   0.12 &   -122 &      14 &      280 &       14 \\
   0.17 &   -143 &      12 &      253 &       12 \\
   0.22 &   -168 &      15 &      277 &       15 \\
   0.27 &   -213 &      18 &      250 &       18 \\
   0.32 &   -162 &      22 &      227 &       22 \\
   0.37 &   -201 &      28 &      299 &       29 \\
   0.42 &   -208 &      37 &      335 &       38 \\
   0.47 &   -119 &      33 &      212 &       34 \\
   0.52 &   -111 &      39 &      211 &       40 \\
   0.57 &   -185 &      57 &      294 &       58 \\
   0.62 &   -169 &      60 &      249 &       61 \\
   0.67 &   -134 &      42 &      214 &       41 \\
   0.72 &   -135 &      39 &      180 &       40 \\
   0.77 &    -46 &      51 &      242 &       51 \\
   0.82 &    -82 &      69 &      324 &       70 \\
   0.87 &     -3 &      71 &      120 &       71 \\
   0.92 &    -77 &      55 &      101 &       56 \\
   0.97 &    -54 &      71 &      293 &       72 \\
\cutinhead{H$\beta$ emission line}
  -0.32 &     55 &      46  &      100 &    47	 \\
  -0.27 &    166 &      38  &      173 &    39	 \\
  -0.22 &    218 &      30  &      145 &    30	 \\
  -0.17 &    187 &      22  &      193 &    23	 \\
  -0.12 &    278 &      18  &      340 &    19	 \\
  -0.07 &    269 &      45  &      508 &    46	 \\
  -0.02 &     11 &      65  &      532 &    67	 \\
   0.02 &    -66 &      51  &      348 &    53	 \\
   0.07 &   -104 &      47  &      224 &    48	 \\
   0.12 &    -95 &      46  &      211 &    47	 \\
   0.17 &   -165 &      51  &       20 &    52	 \\
   0.22 &    -75 &      97  &      184 &    99	 \\
   0.27 &   -298 &      72  &      181 &    73	 \\
\enddata
\end{deluxetable}
\clearpage

The asymmetry in the rotation curve with respect to the nucleus
prevents a good fit to the complete curve using an axisymmetric gas
dynamical model. However, the velocity curve inside $R\la0\farcs3$
shows a clear sign of rotation and one might follow standard practice
of assuming a gas disk. VK00 obtained 6 FOS emission-line spectra in
this region which together with ground-based CTIO emission-line
spectra suggested a thin disk in Keplerian rotation. Under this
assumption, they determined a BH mass of $M_\bullet=1.5\times10^8
M_\odot$ from a combined best-fit to the H$\beta$ and
H$\alpha$+[\ion{N}{2}] gas kinematics (scaling their values to our
adopted distance). The rotation velocities that VK00 obtained for
H$\alpha$+[\ion{N}{2}] with the smallest FOS aperture are compared to
the STIS results in Figure~\ref{fig:gas_modeling}. The data are in
broad agreement, but the STIS data clearly provide a more complete and
detailed view of the gas kinematics within the central arcsecond. In
view of this we have performed a new analysis of the gas kinematics.

We constructed a model for the STIS kinematics using the IDL software
we developed in \citet{ber98}. The model assumes the gas moves on
circular orbits in a thin disk in the galaxy symmetry plane, in the
combined potential of the stars and a possible central BH. Our model
can deal with a general gas surface brightness distribution (e.g. an
observed narrow band image) and takes into account pixel binning, PSF
and slit effects to generate a two-dimensional model spectrum with the
same pixel scale as the observations. Similar to the observations, the
mean model velocity was determined by fitting a Gaussian to each row
of the model spectrum. As the STIS observations use the narrow
0\farcs1 slit, whose width is comparable to the PSF FWHM, we neglect
the velocity offset correction \citep*{van97,mac01,bar01}. A detailed
comparison between our gas modeling software and the completely
independent one used by VK00 shows agreement to better than 5 \kms\
everywhere, for both $V$ and $\sigma$.

For our STIS model we used the $i=60^\circ$ inclination, the
parametric approximation of the disk surface brightness and the
`turbulent' velocity dispersion parameters as derived by VK00. The
stellar potential was determined by deprojecting the axisymmetric MGE
fit to the galaxy surface brightness that was used for the stellar
kinematic modeling and multiplying this with the best fitting stellar
mass-to-light ratio $\Upsilon$. This $\Upsilon$ equals the one used by
VK00 (taking into account the differences in assumed distance).

In Figure~\ref{fig:gas_modeling}, we compare the gas velocities with
models with several different BH masses. The main result is that none
of the models properly fits the gas kinematics. This is not entirely
surprising, given the somewhat disturbed nature of the gas kinematics,
as discussed above. Nonetheless, it is striking to note that the model
with $M_\bullet=2.6\times10^9 M_\odot$, the best-fit value inferred
from the stellar kinematics, does not even come close to fitting the
gas kinematics. It predicts a central rotation curve gradient that is
steeper than what is observed and a a rotation curve amplitude that is
larger than what is observed. The BH mass must be lowered by a factor
of $\sim 7$ to provide a reasonable fit in the central region
$R\lesssim0\farcs3$. The combined STIS, FOS and CTIO data in this
region are best fit by a model with $M_\bullet \approx 3.5\times10^8
M_\odot$. This is somewhat larger than the value inferred by VK00, but
is otherwise in reasonable agreement. A model with no BH predicts a
rotation curve gradient that is considerably more shallow than what is
observed.

\subsection{The BH mass discrepancy}

Discussions of BH demography have recently focused on the scatter in
the relation between BH mass and velocity dispersion. Stellar and gas
kinematical BH mass determinations are consistent with the same
relation and suggest a scatter of $\sim 0.3$ dex \citep[see e.g. the
compilation by][]{tre02}. It is therefore surprising that our analyses
of the stellar and gaseous kinematics in one and the same galaxy have
yielded BH masses that differ by a factor $\sim 7$, i.e., $0.9$
dex. To understand the origin of this discrepancy we proceed by
discussing the potential systematic problems that may have plagued our
analysis.

\subsubsection{Possible problems with the interpretation of the gas kinematics}

There are several possible reasons why our analysis of the gas
kinematics may have yielded an incorrect estimate of the BH mass.

First, it is possible that the underlying assumptions of our model are
correct, but that we have used an incorrect value for the inclination
of the inner gas disk ($i=60^\circ$) or the angle between its
projected major axis and the slit ($\theta=0^\circ$). These angles are
derived from images of the emission-line gas disk at arcsec scales and
larger (see VK00 for discussion). However, the ellipticity of the gas
disk isophotes in the range 0\farcs25--1\farcs0 is
$\epsilon=$0.17--0.37. VK00 interpreted this as thickening of the gas
disk. Alternatively, by allowing the disk to be actually more face-on,
the ellipticity is consistent with inclinations between $\sim
10^\circ$ and $22^\circ$. By constructing models for various
$(i,\theta)$, we find that a model with $i \sim 20^\circ$,
$\theta\sim0^\circ$ and $M_\bullet=2.6\times10^9 M_\odot$ indeed
provides a reasonable and much better fit to the STIS mean gas
velocities at $R\la0.3''$ than a model with $M_\bullet=3.5\times10^8
M_\odot$ (Figure~\ref{fig:gas_modeling}). However, this interpretation
implies that the gas disk is warped (e.g.\ Statler 2001) and that it
cannot reside in the equatorial plane of the galaxy for $R\la1''$ (the
galaxy inclination cannot be lower than $i\la55^\circ$: see Section
4.1). Although the spherical BH potential dominates the galaxy
potential for $R \la 0\farcs65$, the severe warping likely invalidates
the assumption of a thin circular gas disk in equilibrium.

Second, it is possible that the underlying assumptions of our model
are incorrect, and that the gas does not move on circular orbits in an
infinitely thin disk. In our models we have assumed that the observed
velocity dispersion of the gas is `turbulent' and does not contribute
to its hydrostatic support. However, it may be that the gas is better
modeled as individual clouds that move ballistically. In this case the
velocity dispersion would contribute to the hydrostatic support and
the gas would rotate slower than the circular velocity (asymmetric
drift). This would have caused our models to underpredict the BH mass.
If $\sigma / V$ is small, then the upward correction to the BH mass is
small and it can be estimated using approximate equations for
asymmetric drift \citep[e.g.][]{bar01}. However, for IC~1459 the value
of $\sigma / V$ of order unity, and to properly estimate the BH mass
one would have to construct axisymmetric collisionless models for the
gas kinematics.  This is outside the scope of the present
paper. However, one limiting case that can be calculated fairly easily
is to assume that the gas can be modeled through the Jeans equations
as an isotropic spherical distribution of collisionless
cloudlets. This may not be entirely unreasonable, given that the
isophotes of the gas disk become rounder towards the center. We find
that the BH mass must then be $M_\bullet \approx 1.0 \times 10^9
M_\odot$ to reproduce the observed central velocity dispersion of the
gas (this supersedes the somewhat smaller value that we quoted for
this scenario in VK00).

Third, it is possible that our models for the gas kinematics of IC~1459 are flawed at an even more basic level. The gas may not be in
equilibrium. The fact that IC~1459 has a counter-rotating core,
possibly indicative of a recent accretion event, may be of relevance
in this context. Or alternatively, the gas kinematics (mean velocities
{\it and} velocity dispersions) may not be dominated by gravitational
forces. For galaxies in which the central emission-line gas displays a
LINER-type spectrum the gas excitation mechanism might be
shock-ionization \citep[e.g.][]{dop97} as opposed to photo-ionization
(as inferred for Seyferts and Quasars). IC~1459 indeed shows
LINER-type emission and hence shocks may influence the gas kinematics.

IC~1459 was originally identified by us as a good candidate for
thin-disk modeling because of the regular, smooth and prominent
rotation evident in ground-based CTIO spectra. With FOS, spectra could
only be obtained for a few individual apertures (VK00). This yielded
insufficient information to test the hypothesis that the gas is indeed
rotating in a thin disk. The new STIS spectra provide a more complete
view of the gas disk kinematics and clearly show that thin disk models
are oversimplified. The gas does not rotate at all at $R \approx 1''$,
indicating that the true state of the gas must be considerably more
complicated than what we have been able to represent in any of our
models. In hindsight, IC~1459 is a poor galaxy for which to attempt a
gas kinematical determination of the central BH mass. The values that
we have inferred range from $\sim 3.5 \times 10^8 M_\odot$ (thin gas
disk) to $\sim 1.0 \times 10^9 M_\odot$ (spherical distribution of
cloudlets), but we cannot attach strong confidence to either of these
numbers.

\subsubsection{Possible problems with the interpretation of the stellar
kinematics}

The highest spatial resolution stellar kinematical data presented here
are based on a STIS spectrum obtained in 4 orbits of HST time. While
this is a significant exposure time, the S/N of the spectrum is
relatively low (due mostly to the very narrow slit employed).  This
has several consequences. The low S/N, the relatively low spectral
resolution, and the presence of prominent gas emission lines filling
important absorption features, prevent the reliable extraction of
higher-order Gauss--Hermite moments. As a result, we can only obtain
the mean velocity $V$ and velocity dispersion $\sigma$ of the best-fit
Gaussian, with error bars that are considerably larger than those for
the ground-based kinematics. Furthermore, the unresolved source of
featureless non-thermal continuum that dominates the central spectrum
of IC~1459 prevents the measurement of the stellar kinematics inside
0\farcs1. The latter is particularly unfortunate, because the
influence of a BH is always expected to be largest close to the galaxy
center.

\begin{figure}
\epsscale{0.6}
\plotone{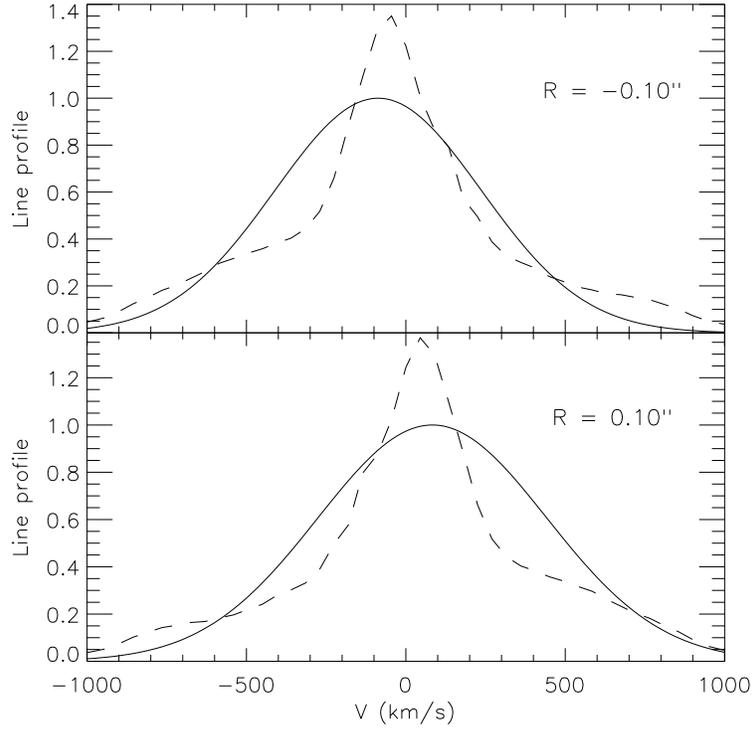}
\caption{The LOSVD predicted by the Schwarzschild modeling with low
regularization ($\Delta=4$, dashed line) at the smallest measured radii
($R=\pm0\farcs10$) within the STIS slit along the major axis, is
compared to the Gaussian measured from the STIS spectrum (solid line).
In both panels the true centered second order moment is $\sigma_{\rm
true}\approx422$ \kms, while the dispersion of the best fitting Gaussian
is $\sigma\approx325$ \kms.
\label{fig:stis_losvd}}
\end{figure}

\begin{figure}
\epsscale{0.6}
\plotone{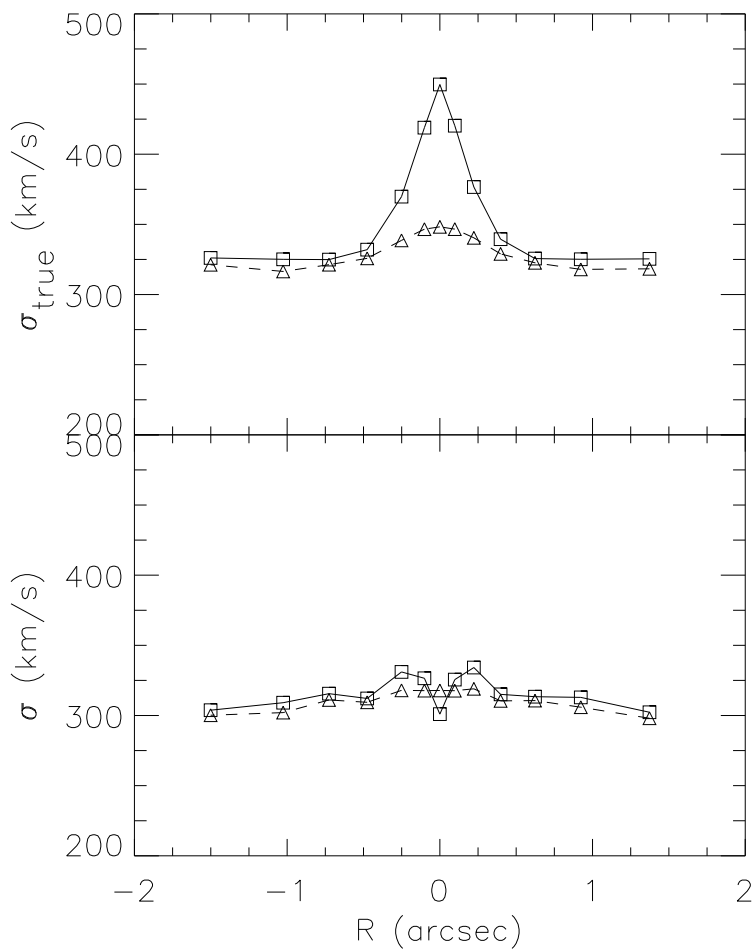}
\caption{\emph{Top Panel:} The true centered second moment $\sigma_{\rm
true}$ predicted by the Schwarzschild modeling, within the STIS slit
along the major axis, for the model with $M_\bullet=3.5\times10^8
M_\odot$ (dashed line with triangles) and with $M_\bullet=2.6\times10^9
M_\odot$ (solid line with open squares). The predictions for the central aperture ($R=0$), where we have no data, is also shown. \emph{Bottom Panel:} Same as in
top panel for the dispersion $\sigma$ of the best fitting
Gaussian.
\label{fig:sigma_true}}
\end{figure}

The $V$ and $\sigma$ profiles derived from the STIS spectrum (for
$R\ga $0\farcs1) do not show any obvious evidence for the presence of
a BH (Figure~\ref{fig:plot_fit_stis}).  While the central velocity
gradient is steeper than the one measured from the ground, there is no
strong central increase in $\sigma$. Also, the value of $\sigma$ at
0\farcs1 is no larger than the central dispersion observed from the
ground. This might suggest a low black hole mass, and indeed, a
dynamical model with the value, derived from the gas kinematics, of
$M_\bullet=3.5\times10^8 M_\odot$, is consistent with the STIS
data. However, our model with the much larger mass of
$M_\bullet=2.6\times10^9 M_\odot$ (inferred from our analysis of all
the stellar kinematical data) is also consistent. The reason for this
surprising fact can be understood by considering the LOSVD predicted
by the model with $M_\bullet=2.6\times10^9 M_\odot$ for the STIS
apertures closest to the center ($R=\pm0\farcs10$;
Fig.~\ref{fig:stis_losvd}). It is strongly non-Gaussian and has very
broad wings, caused by the high-velocity stars near the central
BH. Qualitatively similar profiles, close to the BH, were predicted
with analytic models \citep[e.g.][]{van94a} and have been observed
with HST since \citep[e.g.][]{jos01}. Figure~\ref{fig:sigma_true}
shows that the true centered moment $\sigma_{\rm true}$ displays a
large increase inside the BH sphere of influence ($R_{\rm BH} \approx
0\farcs65$). This rise is absent in $\sigma$ because the best-fitting
Gaussian is quite insensitive to the wings, and as a result the model
can fit the flat $\sigma$ profile even with a large black hole mass.

As a further check of the consistency of our $M_\bullet=2.6\times10^9
M_\odot$ dynamical model with the STIS observations, we convolved a
well-sampled K3 III stellar template with the predicted model LOSVD at
the different apertures along the STIS slit and rebinned this to the
STIS pixel scale. The spectrum thus obtained was processed through the
same procedure, described in Section~\ref{sec:hst_spec}, to extract
the $V$ and $\sigma$. The resulting values agree to within the errors
with the model profiles presented in Figure~\ref{fig:plot_fit_stis}.

\begin{figure}
\epsscale{0.6}
\plotone{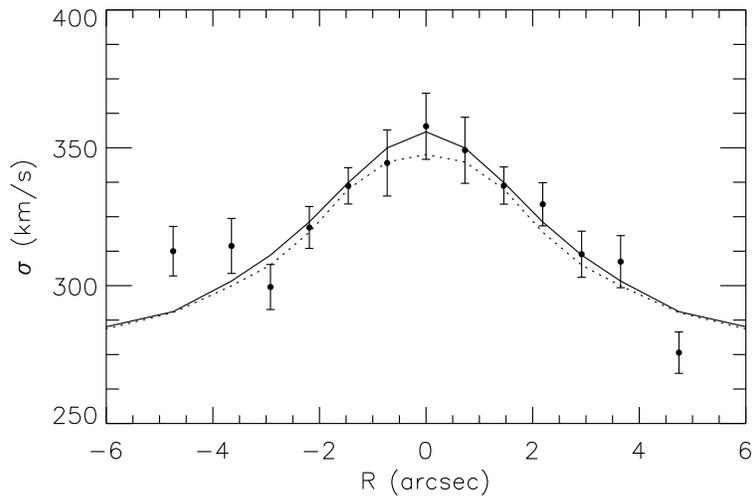}
\caption{Comparison of the velocity dispersion profiles for the
dynamical models with $M_\bullet=3.5\times10^8 M_\odot$ (dotted line)
and with $M_\bullet=2.6\times10^9 M_\odot$ (solid line). This profile
corresponds to the central part of the intermediate axis ground-based
measurements and is also shown in the last column of
Fig.~\ref{fig:plot_fit_major}. The central $\sigma$ predicted by the
model with lower BH mass is always at least one standard deviation below
the measured value, for all observations along five position angles.
Assuming Gaussian errors and considering only the central $\sigma$
measurements, the model with lower BH mass would be excluded at better
than the $3\sigma$ level.
\label{fig:sigma_bh}}
\end{figure}

For the above reasons, our black hole mass estimate is determined
primarily by the ground-based data. Our axisymmetric models take into
account the limited spatial resolution of the data, and they formally
rule out the BH mass $M_\bullet=3.5\times10^8 M_\odot$ that was
suggested by the gas kinematics. This comes about because the central
$\sigma$ measured from the ground is larger than the models can handle
with a black hole mass of less than $\sim 10^9 M_\odot$, while also
fitting the higher order moments. However, Fig.~\ref{fig:sigma_bh}
shows just how small this effect is. The effect is formally quite
significant, but this assumes that there are no uncertainties involved
in the data-model comparison other than Gaussian random errors. This
is obviously an oversimplification. On the one hand, the underlying
assumptions of the models may not be fully correct. For example, IC~1459 could be significantly triaxial, the stars may not yet have
settled into an equilibrium configuration, or line-of-sight
projections may be partially compromised by dust. On the other hand,
there could be systemic problems with the data, e.g., due to continuum
subtraction or template mismatching. Such effects can easily
compromise the data at the few percent level \citep{van94c}. Since the
very central region is affected by a continuum component that is not
present at larger radii, it is quite possible that these errors could
depend on radius. While there are no obvious indications that any of
these issues may be affecting our analysis, it does mean that our BH
mass determination via the stellar kinematics should be treated with
caution.

The accuracy of stellar dynamical BH mass determinations hinges critically on the availability of full LOSVD measurements inside the BH sphere of influence. Unfortunately, we have not been able to secure such a measurements with HST due to the complications introduced by the modest $S/N$, the presence of a non-thermal nuclear point source and of strong gas emission lines. Our BH mass measurement $M_\bullet = (2.6 \pm 1.1) \times 10^9 M_\odot$ therefore rests primarily on $\sim 1.5''$ resolution ground-based spectroscopy. This makes the accuracy of our measurement rather modest, despite the use of three-integral models and the well-defined formal errorbar.

\section{Discussion and conclusions}

We studied the dynamics of the elliptical galaxy IC~1459 using
self-consistent axisymmetric dynamical models based on Schwarzschild's
orbit superposition method. The intrinsic density is obtained by
deprojecting the observed HST/WFPC2 surface density using an MGE
parametrization, after correction for the effects of patchy dust
absorption.  The axisymmetric model provides an excellent fit to all
HST and ground-based kinematical constraints along five slit
positions, while also matching the observed HST photometry. Moreover
we verified that the model velocity histograms indeed reproduce the
observed LOSVDs and not only its moments.

The good match to the observations obtained with an axisymmetric model
provides a test that the counterrotating core in IC~1459 can be
accurately described as an axisymmetric stellar component. We explored
the $\chi^2$ map as a function of the three free parameters of the
problem $(i,\Upsilon,M_\bullet)$ and subsequently analyzed the best
fit solution by adding linear regularization constraints in the
integral space. The observations can only be reproduced if
$\approx0.5$\% of the galaxy stellar mass ($\approx3\times10^9
M_\odot$) counterrotates in the radial range $R\,\lesssim\, 2\arcsec$
on nearly circular orbits in a flattened disk. This counterrotating
component is found to be very well separated from the bulk of the
stars in integral space. The mass estimation of the counterrotating
stellar component is only possible by using dynamical modeling, since
the decoupled component does not show in the surface brightness or
color maps and cannot be isolated from the observed line profiles.

These results could be explained by a scenario in which the stellar
counterrotating core in IC~1459 has an external origin, as evidenced
by its clear isolation in integral space from the bulk of the stars in
the galaxy. For example, the core may have formed by acquisition of a
gas-rich component that settled into a disk in the galaxy potential
and later turned into stars. In this scenario it is likely that the
acquisition must have happened long ago. The current stellar core
counterrotates not only with respect to the outer stellar body but
also with respect to the nuclear gas-disk of IC~1459, which must have
formed later. Numerical simulations of gas accretion by a BH that
leads to disk formation and BH fueling were made by e.g. \citet{bek00}
However, other formation scenarios are also possible. For example,
\citet{hol00} showed that, in the presence of a central BH, a rapidly
rotating counterrotating core can also be produced by acquisition of a
dense stellar satellite.

IC~1459 has a rotating gas component near its center. This offers the opportunity to estimate the BH mass independently from either gaseous or stellar kinematics. For both we have available high spatial resolution long-slit kinematics from STIS, as well as ground-based kinematics along multiple position angles. Unfortunately, we have found that neither the gaseous nor the stellar kinematics yield a very stringent determination of the BH mass. The main problem for the gas kinematical modeling is that there is evidence that the gas motions are disturbed, possibly due to non-gravitational forces acting on the gas. The main problem for the stellar kinematical modeling is that we were unable to obtain an HST measurement of the stellar LOSVD inside the BH sphere of influence, due to the modest $S/N$ of our STIS spectrum and due to the presence of the non-thermal nuclear point source of IC~1459 and prominent gas emission lines. Consequently, the stellar dynamically inferred BH mass hinges almost exclusively on the low spatial-resolution ground-based data. These complications may explain why we find rather discrepant BH masses with the different methods. The gas kinematics suggests that $M_\bullet = 3.5 \times 10^8 M_\odot$ if the gas is assumed to rotate at the circular velocity in a thin disk. Alternatively, the central velocity dispersion of the gas implies that $M_\bullet = 1.0 \times 10^9 M_\odot$ if the gas is modeled as an isotropic spherical distribution of ballistic cloudlets. The stellar kinematics suggest that $M_\bullet = (2.6 \pm 1.1) \times 10^9 M_\odot$. These different estimates bracket the value $M_\bullet = (1.1\pm0.3)\times10^9 M_\odot$ predicted by the $M_\bullet$-$\sigma$ relation \citep{fer00,geb00} as given by Tremaine et al.\ (2002).

To our knowledge, IC~1459 is only the second galaxy (Pinkney et al.\
2000 modeled NGC 4697) for which a BH mass determination has been
attempted with HST observations of both gaseous and stellar
kinematics. Unfortunately, as it turned out, IC~1459 may not been the
best galaxy to attempt such a comparison. However, it certainly is
important to perform such comparisons. They provide insight into the
reliability of BH mass determinations, which is necessary to properly
understand the correlation between the BH mass correlation and other
global galaxy parameters (Tremaine et al.\ 2002 and the refs
therein). This is particularly relevant if one wants to understand the
scatter in such relations and their implications for our understanding
of the galaxy formation process.

With the availability of 2D spectroscopic detectors (e.g. {\tt SAURON}
on WHT, {\tt VIMOS} on VLT, {\tt GMOS} on Gemini) it is possible to
obtain fully two-dimensional distributions of LOSVDs and
line-strengths of kinematically decoupled cores
\citep[e.g.][]{dav01,dez02}.  These observations will provide a much
larger number of kinematical constraints than can be obtained from a
small number of slit positions, which results in dynamical models that
better constrain the model parameters \citep[e.g.][]{ver02},
especially if they resolve the radius of influence of the BH. In turn,
this will improve the reliability of internal dynamical studies such
as that presented in this paper.

\acknowledgements
Support for proposal 7352 was provided by NASA through a grant from the Space Telescope Science Institute, which is operated by the Association of Universities for Research in Astronomy, Inc., under NASA contract NAS 5-26555.
This work was initiated during a visit of MC to Leiden Observatory. We thank Karl Gebhardt for useful comments on an earlier draft of this manuscript.


\appendix

\section{MGE potential calculation}
\label{sec:potential}

In this Appendix, we discuss the techniques that were used to efficiently and accurately  evaluate the MGE potential. The MGE surface brightness defined in equation~(\ref{eq:mge_surf}) can be deprojected  analytically \citep{mon92} to derive an axisymmetric intrinsic luminosity density consistent with the observations. Using the equations in \citet{cap02} the intrinsic luminosity density can be written as
\begin{equation}
\rho(R,z) = \sum_{j=1}^N
\frac{L_j}{(\sigma_j\,\sqrt{2\pi})^3 q_j} \exp
\left[
    -\frac{1}{2\sigma_j^2}
    \left(R^2+\frac{z^2}{q_j^2} \right)
\right],
\label{eq:dens}
\end{equation}
where $N$ is the number of Gaussians required in the expansion, $L_j$ is the Gaussian total luminosity, $\sigma_j$ the corresponding dispersion and the intrinsic axial ratio $q_j=\sqrt{q'^2_j-\cos^2 i}/\sin i$, where $i$ is the galaxy inclination.

The general form for the potential that corresponds to an MGE density profile is
\begin{equation}
\Psi(R,z) = G \Upsilon\sqrt{2/\pi}\; \sum_{j=1}^N
{\frac{L_j\, {\mathcal Q}_j(R,z)}{\sigma_j}}
\label{eq:mge_potential}
\end{equation}
\citep{ems94}, with
\begin{equation}
{\mathcal Q}_j(R,z) = \int_0^1 {\frac{{\exp \left\{ - \frac{{T^2 }}
{{2\sigma _j^2 }}\left[ {R^2  + \frac{{z^2 }}{{1 - (1 - q_j^2 )T^2 }}}
\right] \right\}}}{{\sqrt {1 - (1 - q_j^2 )T^2 } }} \ud T}.
\label{eq:integral}
\end{equation}

Since the Gaussians of an MGE parametrization sample many orders of magnitude in radius, it is worthwhile to check if it is necessary to do the actual integration in equation~(\ref{eq:integral}) or whether a limiting expression can be used instead. We desire a maximum relative error of $\epsilon\,\lesssim\,10^{-4}$ for the potential. This is at least two orders of magnitude more accurate than what can generally be achieved for photometric measurements.

An approximate expression for the potential at large radii can be found using a multipole expansion of the potential \citep[e.g.][]{bin87}. Terminating the expansion at the first term, ${\mathcal Q}_j(R,z)$ in equation~(\ref{eq:integral}) equals
\begin{equation}
{\mathcal Q}_j(R,z) = \frac{\sigma_j\,\sqrt{\frac{\pi}{2}}}{\sqrt{R^2+z^2}},
\end{equation}
and the potential reduces to the potential of a point mass with mass $\mathcal{M}=\Upsilon L_j$. Even in the thin disk limit, a maximum relative error $\epsilon\,\lesssim\,10^{-4}$ is obtained outside the sphere $100\,\sigma_j \la \sqrt{R^2+z^2}$. In the case of positive Gaussians this is a conservative estimate for the total relative error in the potential, since it applies only to one Gaussian, while in practice not all Gaussians in the expansion will be in the least favorable condition at the same time. This error limitation is generally not true when the MGE is composed by both positive and negative Gaussians, due to cancellation effects.

An approximate expression for the potential at small radii can be obtained by calculating a Taylor expansion of the $\exp(x)$ term in equation~(\ref{eq:integral}) around $x=0$. The first term in this expansion equals
\begin{equation}
{\mathcal Q}_j(R,z) = \frac{\arcsin \left({\sqrt{1 - q_j^2}}\right)}{{\sqrt{1 -
q_j^2}}}.
\label{eq:central_potential}
\end{equation}
This corresponds to replacing the potential by its central analytic value  $\Psi(0,0)$. Assuming a lower limit on the axial ratio $0.1<q_j$ a maximum relative error $\epsilon\,\lesssim\,10^{-4}$ is obtained inside the sphere $\sqrt{R^2+z^2} \la\,\sigma_j/171$.

An example of the usefulness of this approximation is given by the present application. Our orbit grid spans the radial range $0\farcs05<R<300\arcsec$; a comparison with Table~\ref{tab:ic1459_mge} therefore shows that these  approximations are used often and only a small number of integrations has to be carried out. A related advantage of this is that the regions where the approximations are used are precisely those where the integrand is badly behaved and the numerical integration may become inaccurate.

Similar considerations are valid for the two components of the acceleration in the meridional plane, which are required by the orbit integration.

When the integral in equation~(\ref{eq:integral}) has to be evaluated, we use a globally adaptive, recursive Gauss-Kronrod integration scheme \citep*{fav91}. The use of such an algorithm is very beneficial at small and large distances from the galaxy center (but before the switch to the limiting regimes), where the integrand of equation~ (\ref{eq:integral}) shows a sharp peak close to $T=1$ or to $T=0$ respectively.  In these cases the integrand is evaluated more densely around the peaks.

\section{Integration of a Gaussian on a grid}
\label{sec:integration}

Axisymmetric Schwarzschild modeling is aimed at finding the linear combination of orbits that reproduces the observed kinematics at various apertures, the intrinsic density $\rho(R,z)$ on the meridional plane and, to limit the effect of discretization, also the surface density $\Sigma(x,y)$ on the projected plane. In general, the comparison with the density requires a double or triple integration. However, these computations can be simplified for an MGE parametrization. We discuss the different cases below. We always assume that the Gaussians have a maximum value of one, which means that intrinsic quantities have to be multiplied by $L_j/[(\sigma_j\,\sqrt{2\pi})^3 q_j]$ and projected quantities by $L_j/(2\pi\, \sigma_j^2 q'_j)$ \citep[see][for details]{cap02}. In the following equations, the index has been dropped from the $\sigma_j$ and $q_j$ variables and $q$ has to be substituted by $q'$ when dealing with projected plane quantities.

\subsection{Polar grid in the meridional plane}

\begin{figure}[t]
\epsscale{0.6}
\plotone{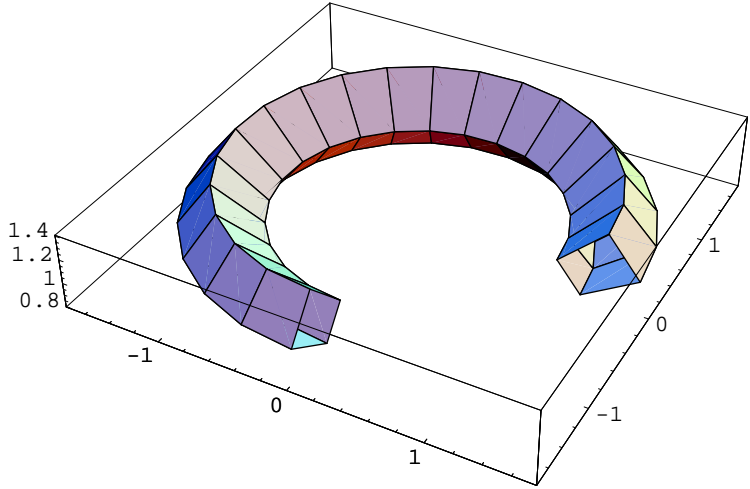}
\caption{One of the elements of the meridional plane grid. Equation~ \ref{eq:merid_integ} represents the mass of an axisymmetric Gaussian contained within this region. One quadrant has been removed from the element for display purposes. \label{fig:polar_integral_meridional}}
\end{figure}

We integrate an axisymmetric three dimensional Gaussian defined by $\rho(R,z)$ in equation~(\ref{eq:dens}) on the ``torus'' $D$ of Figure~\ref{fig:polar_integral_meridional} around the galaxy symmetry axis, which is delimited by the intervals $[r_0,r_1]$ and $[\theta_0,\theta_1]$ in spherical coordinates ($\theta=0$ on the symmetry axis). The integrated mass is given in spherical coordinates by
\begin{equation}
\int_D\rho\, =\,
2\pi \int_{\theta_0}^{\theta_1} \int_{r_0}^{r_1}
        e^{-(r/c(\theta))^2}\,r^2\,\sin\theta  \ud r\, \ud\theta
\end{equation}
where $c(\theta)=2\,q\,\sigma\,/\sqrt{1+q^2+(1-q^2)\cos2\theta}$. The
innermost integral can be performed analytically and the integral
becomes
\begin{equation}
\pi\, \int_{\theta_0}^{\theta_1}
        \bigg\{
            \frac{c(\theta)\sqrt{\pi}}{2}\,
            \left[
                {\rm erf}\left(\frac{r_1}{c(\theta)}\right) -
                {\rm erf}\left(\frac{r_0}{c(\theta)}\right)
            \right] +
            r_0\,e^{-(r_0/c(\theta))^2}
            -\,r_1\,e^{-(r_1/c(\theta))^2}
        \bigg\}\,
        c(\theta)^2\,\sin\theta  \ud\theta
\label{eq:merid_integ}
\end{equation}

\subsection{Polar grid in the projected plane}

In the case of the integration of a two dimensional Gaussian on a polar cell $D$, delimited by the intervals $[r_0,r_1]$ and $[\theta_0,\theta_1]$ in polar coordinates, we have to compute the integral
\begin{equation}
\int_D\Sigma\, =\,
\int_{\theta_0}^{\theta_1} \int_{r_0}^{r_1}
e^{-(r/c(\theta))^2}\, r\,  \ud r\, \ud\theta,
\end{equation}
where $\theta=0$ on the Gaussian minor axis, which gives
\begin{equation}
\int_{\theta_0}^{\theta_1} \frac{c(\theta)^2}{2}
    \left[e^{-(r_0/c(\theta))^2} - e^{-(r_1/c(\theta))^2}\right]
    \ud\theta.
\end{equation}

\subsection{Cartesian grid on the projected plane}

We compute the integral of a two-dimensional Gaussian on a Cartesian grid cell $D$ delimited by the intervals $[x_0,x_1]$ and $[y_0,y_1]$. Assuming the cell is parallel to the Gaussian major axis, the integral is given by
\begin{equation}
\int_D\Sigma\, =\,
\frac{\pi \,q\,\sigma^2}{2}
    \left[
        {\rm erf}\left(\frac{x_0}{\sigma\sqrt{2}}\right) -
        {\rm erf}\left(\frac{x_1}{\sigma\sqrt{2}}\right)
    \right]
  \left[
        {\rm erf}\left(\frac{y_0}{q\,\sigma\sqrt{2}}\right) -
        {\rm erf}\left(\frac{y_1}{q\,\sigma\sqrt{2}}\right)
    \right].
\end{equation}
If the Gaussian major axis makes an angle $\theta$ with the $x$ axis, the double integral on the grid cannot be evaluated explicitly, but can still be expressed as a single integral in one of the coordinates as follows
\begin{equation}
\frac{q\,\sigma\sqrt{\pi}}{p}\!
\int_{x_0}^{x_1}\!\!
    \left[
        {\rm erf}\!\left(\frac{f_0(x)}{2 p q \sigma}\right)\!\! -\!
        {\rm erf}\!\left(\frac{f_1(x)}{2 p q \sigma}\right)
    \right]\! e^{-[x/(p\,\sigma)]^2}  \ud x,
\end{equation}
with $p=\sqrt{1+q^2+(1-q^2)\cos2\theta}$, and we defined
\begin{equation}
f_k(x) = (1-q^2)\,x\,\sin2\theta -p^2\,y_k \qquad {\rm for }\qquad k=0,1.
\end{equation}

\section{Hermite polynomials}

We tabulate the first seven Hermite orthogonal polynomials, following the normalization of \citet{van93}. They are written in Horner form for efficient and stable evaluation. The first five are also given in equation~(A5) of that paper.
\begin{eqnarray}
 H_0(y) & = & 1, \quad
 H_1(y) = {\sqrt{2}}\,y, \quad
 H_2(y) = \frac{2\,y^2 - 1}{{\sqrt{2}}}, \quad
 H_3(y) = \frac{y\,\left(2\,y^2  -3 \right) }{{\sqrt{3}}}, \nonumber\\
 H_4(y) & = & \frac{y^2\,\left(4\,y^2 -12 \right) + 3}{2\,{\sqrt{6}}}, \quad
 H_5(y) = \frac{y\,\left[ y^2\,\left( 4\,y^2  -20 \right) + 15 \right] }
                    {2\,{\sqrt{15}}}, \nonumber\\
 H_6(y) & = & \frac{ y^2\,\left[ y^2\,\left( 8\,y^2 -60 \right) + 90 \right] -15}
                {12\,{\sqrt{5}}}.
\end{eqnarray}


\end{document}